\documentclass{article}

\usepackage{microtype}
\usepackage{graphicx}
\usepackage{booktabs} 
\usepackage{array}
\usepackage[version=4,arrows=pgf-filled,textfontname=sffamily,mathfontname=mathsf]{mhchem}
\usepackage[backref=page]{hyperref}

\usepackage{amsmath,amsfonts,bm}

\def\eqref#1{equation~\ref{#1}}

\def\1{\bm{1}}

\def\rmC{{\mathbf{C}}}

\def\rmR{{\mathbf{R}}}

\def\rmX{{\mathbf{X}}}

\def\va{{\bm{a}}}
\def\vb{{\bm{b}}}

\def\vf{{\bm{f}}}

\def\vk{{\bm{k}}}

\def\vr{{\bm{r}}}

\def\vt{{\bm{t}}}

\def\vv{{\bm{v}}}

\def\vx{{\bm{x}}}

\def\vz{{\bm{z}}}

\def\mL{{\bm{L}}}

\DeclareMathAlphabet{\mathsfit}{\encodingdefault}{\sfdefault}{m}{sl}
\SetMathAlphabet{\mathsfit}{bold}{\encodingdefault}{\sfdefault}{bx}{n}

\def\gA{{\mathcal{A}}}

\def\gF{{\mathcal{F}}}

\def\gK{{\mathcal{K}}}
\def\gL{{\mathcal{L}}}
\def\gM{{\mathcal{M}}}
\def\gN{{\mathcal{N}}}

\def\gP{{\mathcal{P}}}
\def\gQ{{\mathcal{Q}}}
\def\gR{{\mathcal{R}}}

\def\gT{{\mathcal{T}}}

\def\gV{{\mathcal{V}}}

\def\gX{{\mathcal{X}}}
\def\gY{{\mathcal{Y}}}

\newcommand{\R}{\mathbb{R}}

\usepackage[accepted]{icml2024}

\usepackage{amsmath}
\usepackage{amssymb}
\usepackage{mathtools}
\usepackage{amsthm}
\usepackage{booktabs}
\usepackage{multirow}
\usepackage{pifont}
\usepackage{enumitem}

\newcommand{\cmark}{\ding{51}}%
\newcommand{\xmark}{\ding{55}}%

\usepackage[capitalize,noabbrev]{cleveref}

\theoremstyle{plain}

\theoremstyle{definition}

\theoremstyle{remark}

\usepackage{lipsum}
\usepackage{wrapfig}
\usepackage{color}
\usepackage{commath}
\usepackage{caption,subcaption}

\definecolor{myblue}{RGB}{0, 0, 255} 
\definecolor{mydarkblue}{RGB}{0, 0, 139} 
\hypersetup{
    colorlinks=true,
    linkcolor=mydarkblue,
    filecolor=mydarkblue,
    citecolor=mydarkblue,
    urlcolor=mydarkblue
}

\usepackage[textsize=tiny]{todonotes}

\icmltitlerunning{Gaussian Plane-Wave Neural Operator for Electron Density Estimation}

\begin{document}

\twocolumn[
\icmltitle{Gaussian Plane-Wave Neural Operator for Electron Density Estimation}



\icmlsetsymbol{equal}{*}

\begin{icmlauthorlist}
\icmlauthor{Seongsu Kim}{postech}
\icmlauthor{Sungsoo Ahn}{postech}
\end{icmlauthorlist}

\icmlaffiliation{postech}{Pohang University of Science and Technology
(POSTECH), Pohang, South Korea}

\icmlcorrespondingauthor{Seongsu~Kim}{seongsukim@postech.ac.kr}
\icmlcorrespondingauthor{Sungsoo~Ahn}{\mbox{sungsoo.ahn}@postech.ac.kr}

\icmlkeywords{Machine Learning, ICML}

\vskip 0.3in
]
\printAffiliationsAndNotice{}  


\begin{abstract}
This work studies machine learning for electron density prediction, which is fundamental for understanding chemical systems and density functional theory (DFT) simulations. To this end, we introduce the Gaussian plane-wave neural operator (GPWNO), which operates in the infinite-dimensional functional space using the plane-wave and Gaussian-type orbital bases, widely recognized in the context of DFT. In particular, both high- and low-frequency components of the density can be effectively represented due to the complementary nature of the two bases. Extensive experiments on QM9, MD, and material project datasets demonstrate GPWNO's superior performance over ten baselines.
\end{abstract}

\section{Introduction}
The study of electron density is fundamental for understanding the properties and behaviors of chemical systems. Its estimation is a crucial component of Kohn-Sham density functional theory \citep[DFT]{kohn1965self} for simulations like ground-state energy estimation with applications to the development of new battery cathodes \citep{kondrakov2017charge}, solar cell materials \citep{chang2021lead} or drug design \citep{adekoya2022application}. However, the computational complexity of the DFT is cubic in the system size and is often prohibitively expensive for real-world problems.

Machine learning has shown promise in predicting electron density with less computational complexity. Early works investigated methods based on kernel ridge regression \citep{grisafi2018transferable, bogojeski2018efficient} and Gaussian process-based regression \citep{grisafi2018transferable, fabrizio2019electron}. Recently, deep learning methods have been increasingly investigated due to their remarkable capability to capture complex relationships within data. \citet{sinitskiy2018deep} utilized a voxel-based 3D convolutional net with a U-Net architecture to predict density at a voxel level. \citet{jorgensen2020deepdft} and \citet{jorgensen2022equivariant} aimed to predict the electron density using graph neural networks, where the graph consists of interacting atom vertices and a special query point vertex for which the density is predicted. 

More recently, \citet{cheng2023equivariant} tackled the problem with neural operator, an attractive paradigm that can map the given molecule into a function space of electron density whose output can be evaluated at an arbitrary point. In particular, they proposed to predict the density using atom-centered Gaussian-type orbital \citep[GTO]{hehre1972self}.



In particular, exploration of the plane-wave (PW) basis, a cornerstone in DFT solvers like VASP~\citep{hafner2008ab}, Quantum Espresso~\citep{giannozzi2017advanced}, and ABINIT~\citep{gonze2009abinit}, remains absent within the domain of neural operators. The PW basis comprises periodic wavefunctions that meet the periodic boundary condition of the molecule. It is an appealing approach due to its ease of manipulation via fast Fourier transformation and its ability to effectively capture low-frequency signals such as those associated with long-range interactions, e.g., Coloumb potentials. The most similar work is the Fourier neural operator \citep[FNO]{li2020fourier} baseline considered by \citet{cheng2023equivariant}, which uses the Fourier basis consisting of periodic functions. The difference is how the Fourier basis does not respect the periodic boundary condition while the PW basis does.

\setlength{\belowcaptionskip}{-1em}
\begin{figure*}[t]
\begin{minipage}{0.34\linewidth}
  \subfloat{\includegraphics[width=\linewidth]{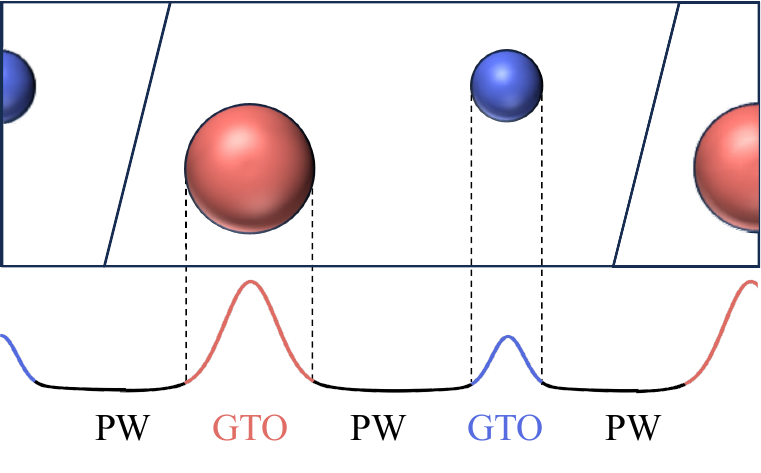}}  
\end{minipage}
\hfill
\begin{minipage}{0.64\linewidth}
  \subfloat{\includegraphics[width=\linewidth]{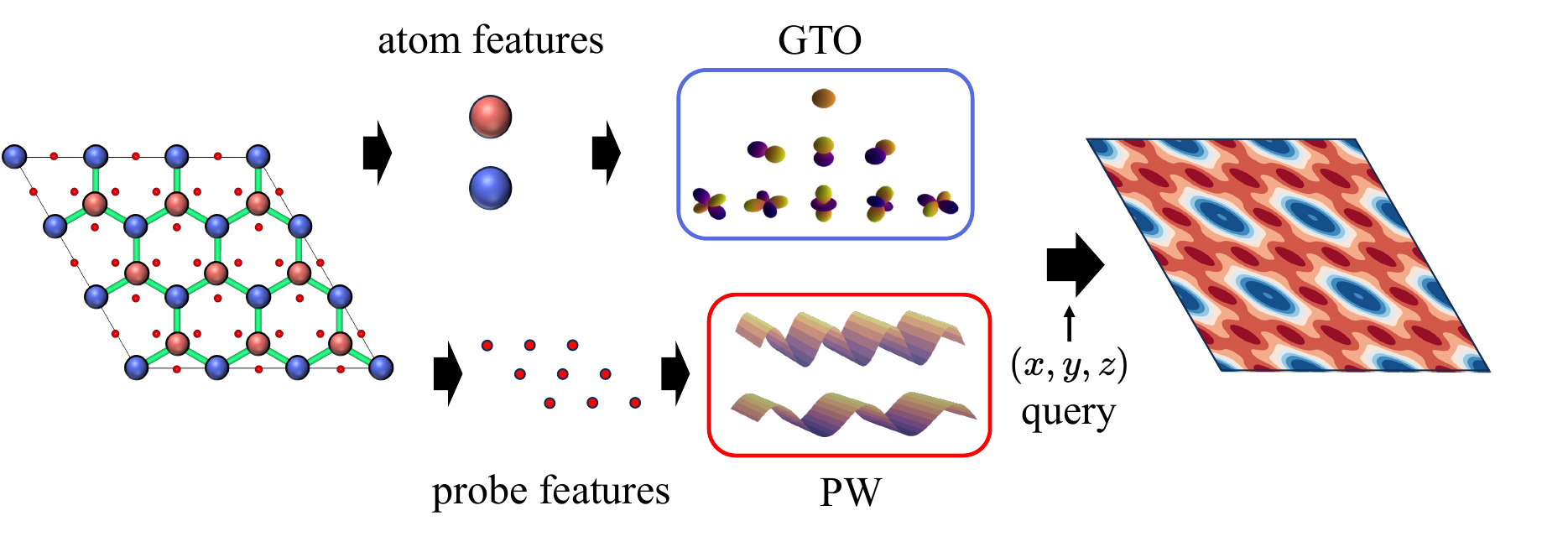}}
  \end{minipage}
\vspace{-2mm}\caption{\textbf{Illustration of Gaussian plane wave neural operator (GPWNO).} \textbf{(left)} Our GPWNO decomposes the electron density estimation into two regions: high-frequency and fast-decaying regions for GTO basis and low-frequency and slow-decaying regions for the PW basis. \textbf{(right)} Our GPWNO makes two predictions based on the GTO and the PW bases. The PW-based prediction is from the lattice-based discretization via the probe nodes, while the GTO-based prediction is from the atom-wise message-passing scheme. Final output is evaluated at an arbitrary query point.}
\label{fig:overview}
\end{figure*}
\setlength{\belowcaptionskip}{0em}


Motivated by this lack of exploration, we aim to predict the electron density using neural operators that operate on the PW basis.
However, using a moderately-sized PW basis, e.g., with $10^{3}$ elements, alone cannot accurately express the electron density, which is in agreement with the poor performance of FNO reported by \citet{cheng2023equivariant}. While DFT solvers overcome this by increasing the size, e.g., up to $8.9\times10^{7}$, the neural operators are not scalable to such high frequencies. Our focus is to enjoy the benefits of PW basis while keeping the basis size realistic for neural operators. Prior works on DFT, e.g., pseudopotential~\citep{harrison1966pseudopotentials,bachelet1982pseudopotentials}, linearized augmented plane-wave~\citep{koelling1975use}, projector augmented wave \citep{blochl1994projector}, and mixed density fitting \citep{sun2017gaussian}, hint that it is possible to reduce the required basis size by decoupling vicinity of atoms from the rest of the regions and mixing local orbitals with PWs.

\textbf{Contribution.} In this work, we propose Gaussian plane-wave neural operator (GPWNO) for estimating the electron density. Our main idea is to decompose the electron density into local atom-wise Gaussian-type orbital (GTO) and global plane-wave (PW) bases, where the bases complement each other for the expression of high- and low-frequency signals. In addition, our design choice allows prediction that is both SE(3)-equivariant and satisfies the periodic boundary condition. We provide an overview of GPWNO in \cref{fig:overview}.

Our main component is the PW-based electron density prediction, which is based on PW neural operator layers with the convolution parameterized in the reciprocal space of the molecule, i.e., PW coefficients. To this end, we discretize the signal via probe nodes regularly placed along the lattice structure and apply discrete Fourier transformations to compute the convolution. Finally, the density at query nodes is evaluated via message passing. Our main idea on the lattice-based discretization, which allows convolution of the signals in the reciprocal space of molecules while satisfying the periodic boundary condition. 


Our next component is the GTO-based electron density prediction, which can be defined as a neural operator in the graphon space  \citep{cheng2023equivariant}. We also introduce high-frequency masking scheme that allows the PW basis to concentrate on regions outside the vicinity of atoms. Our main idea here is to exploit the nature of GTO complementing the PW basis, which was inspired by DFT solvers like Gaussian plane-wave methods \citep{lippert1997hybrid}, LAPW with local orbitals \citep{koelling1975use}, and mixed density fitting \citep{sun2017gaussian}.

We empirically verify the effectiveness of our methods on three datasets: QM9 \citep{ruddigkeit2012enumeration, ramakrishnan2014quantum, Jørgensen_Bhowmik_2022}, MD \citep{brockherde2017bypassing, bogojeski2020quantum} datasets. We additionally benchmark our algorithm on the newly curated datasets from the material project \citep[MP]{jain2013commentary,shen2021representationindependent}, categorized by seven different crystal family types, e.g., the crystal being triclinic and hexagonal. We show that our algorithm consistently outperforms the prior works by a large margin. We validate the effectiveness of our algorithmic components through an ablation study. 

In summary, our contribution is twofold:
\begin{itemize}[topsep=0pt]
    \item We propose GPWNO for electron density prediction based on the mixture of GTO and PW bases which complement each other for the expression of high- and low-frequency components in the density. We design the framework to satisfy SE(3)-equivariance and periodic boundary conditions.
    \vspace{-3.3mm}\item We empirically validate our GPWNO for the QM9, MD, and newly curated MP datasets with comparison against ten baselines. We also empirically verify the effectiveness of each component of our algorithm through ablation studies.
\end{itemize} 
\section{Related Works}

\textbf{Machine learning for electron density estimation.} Initially, researchers have resorted to statistical approaches such as kernel ridge regression \citep{grisafi2018transferable, bogojeski2018efficient} and symmetry-adapted Gaussian processes~\citep{grisafi2018transferable, fabrizio2019electron} for electron density estimation. Recent advance in deep learning have opened new avenues for modeling these functions. For example, \citet{sinitskiy2018deep} employed a voxel-based 3D U-Net convolutional network architecture, to predict electron densities at the voxel level. 

Furthermore, \citet{jorgensen2020deepdft} and \citet{jorgensen2022equivariant} approached electron density prediction using neural message passing on graphs. \citet{koker2023higher} advanced this field by utilizing higher-order equivariant features for more expressive representations. Finally, \citet{cheng2023equivariant} embraced the idea of coefficient learning within the multicentric approximation framework and developed an equivariant graph neural network (GNN) based on tensor products for predicting density spectra.

\textbf{Neural operator.} Neural operators \citep[NOs]{li2020neural, li2020fourier, kovachki2023neural, tran2023factorized} form a paradigm to map data into infinite-dimensional function spaces where these spaces and the data share a common domain, e.g., from 3D vehicle geometries to pressure on car surfaces \citep{li2023geometry}, from molecular systems to electron densities \citep{cheng2023equivariant} and from seismic wavefield to probability to P- and S- wave arrivals \citep{sun2023phase}. The input function for neural operators can be represented with any level of discretization, grid, resolution, or mesh. The output function, in turn, can be evaluated at any arbitrary point.





Initially, the graph neural operator \citep[GNO]{li2020neural} proposed to integrate kernel operation within graph structures, drawing similarities to the message-passing networks. Next, the Fourier neural operator \citep[FNO]{li2020fourier} aims to parameterize and learn operators within the Fourier domain. This is done by discretizing the input domain and applying the fast Fourier transform (FFT). To improve the scalability of FNO, factorized-FNO \citep[F-FNO]{tran2023factorized}, employs Fourier factorization, learning features in the Fourier space independently across each dimension. Finally, to overcome the limits of FNO being restricted to regular grids for using FFT operations, the geometry-informed neural operator~\citep[GINO]{li2023geometry} maps irregular grids into a latent regular grid and then applies FFT.
\section{Preliminary}
\subsection{Problem Definition}
Here, we formulate the electron density prediction problem as learning a mapping from a three-dimensional Cartesian coordinate to a real number, given a molecule. We consider \textit{periodic} molecules, where the periodicity is defined by the lattice structure. We also discuss extending our framework to aperiodic molecules.

To be specific, we represent a molecule of $N$ atoms as $\mathcal{M}=(\mathcal{X}, \mathcal{A}, \mL)$ with atom positions $\mathcal{X}$, atomic features $\mathcal{A}$, and unit cell lattice $\mL$. The atomic positions $\mathcal{X}=\{\vx_{u}: u\in\mathcal{V}\}$ and the atom features $\mathcal{A}=\{\vz_{u}: u\in\mathcal{V}\}$ where we use vertices $\mathcal{V}$ to index the atoms. Finally, we let the unit cell lattice $\mL = [\va_{1}, \va_{2}, \va_{3}]^{\top}$ describe the periodicity where each $\va_{i}$ represents a lattice vector. 

Given the molecule $\mathcal{M}=(\mathcal{X}, \mathcal{A}, \mL)$, the infinite molecular representation $(\bar{\mathcal{X}}, \bar{\mathcal{A}})$ with atom positions $\bar{\mathcal{X}}$ and atom features $\bar{\mathcal{A}}$ is defined as follows:
\begin{align}
    \bar{\mathcal{X}} &= \bigg\{\vx_{\vt, u} \bigg| \vx_{\vt, u} = \vx_{u} + \sum_{d=1}^{3}t_{d} \va_{d}, \vt\in\mathbb{Z}^{3}, u\in \mathcal{V}\bigg\},
\end{align}
\begin{align}
    \bar{\mathcal{A}} &= \{\vz_{\vt, u} | \vz_{\vt, u} = \vz_{u}, \vt\in\mathbb{Z}^{3}, u\in \mathcal{V}\},
\end{align}
where $\vt=[t_{1}, t_{2}, t_{3}]\in\mathbb{Z}^{3}$ indicates the change of unit cell over the infinitely repeating periodic patterns.



Consequently, the true electron density function $\rho:\mathbb{R}^{3}\rightarrow \mathbb{R}$ also satisfies the following \textit{periodic boundary condition}:
\begin{equation}\label{eq:pbc}
    \rho(\vx) = \rho\bigg(\vx + \sum_{d=1}^{3}t_{d} \va_{d}\bigg),
\end{equation}
where  $\vt=[t_{1}, t_{2}, t_{3}]\in\mathbb{Z}^{3}$ indicates the change of unit cell over the infinitely repeating periodic patterns.

\textbf{Extension to aperiodic molecules.} For aperiodic molecules, we follow the common practice of formulating them as periodic molecules with a sufficiently large unit cell lattice~$\mL$ with zero-padded boundaries. Examples include DFT solvers \citep{blochl1995electrostatic,hafner2008ab,ulian2013comparison} and graph neural networks~\citep{kosmala2023ewaldbased}. 

\subsection{Plane-Wave Basis}
The plane-wave (PW) basis is widely used in electronic structure computation of solid-state systems \citep{martin_2004,kittel2005introduction} because it satisfies the periodic boundary condition by construction and is easily manipulated through fast Fourier transformations. We consider the truncated version of the (infinite) PW basis determined by a cutoff in the maximum frequency (or energy), as commonly employed by DFT solvers and Fourier neural operators.

Formally, given a lattice $\mL = [\va_{1}, \va_{2}, \va_{3}]^{\top}$, the wavevectors are constructed from reciprocal primitive vectors $(\vb_{1}, \vb_{2}, \vb_{3})$ where $\vb_{1}$ is defined by $(\va_{1}, \va_{2}, \va_{3})$ as:
\begin{equation}
    \vb_{1} = 2\pi\frac{\va_{2}\times \va_{3}}{\va_{1}\cdot (\va_{2}\times \va_{3})}
\end{equation}
and $\vb_{2}$ and $\vb_{3}$ are defined similarly by $(\va_{2}, \va_{3}, \va_{1})$ and $(\va_{3}, \va_{1}, \va_{2})$, respectively. Note the dual relationship between the lattice vectors and the primitive vectors, e.g., $\va_{1}\cdot \vb_{1} = 2\pi$ and $\va_{1}\cdot \vb_{2} = 0$.

Finally, the PW basis element $\phi_{\bm{\lambda}}(\vx)$ indexed by a tuple of integers $\bm{\lambda}=[\lambda_{1},\lambda_{2},\lambda_{3}]^{\top}\in\mathbb{Z}^{3}$ is defined as follows:
\begin{equation}\label{eq:pwbasis}
    \phi_{\bm{\lambda}}(\vx) = \exp\bigg(j\sum_{d=1}^{3}\lambda_{d}\vb_{d}\cdot \vx\bigg),
\end{equation}
where $j$ is the imaginary number. It should be noted that the plane wave basis element for any $\bm{\lambda}=[\lambda_{1},\lambda_{2},\lambda_{3}]^{\top}\in\mathbb{Z}^{3}$ satisfies the periodic boundary condition, i.e. \cref{eq:pbc}. We also point out how the PW basis was originally designed for the wavefunctions to solve Schr\"odinger's equation \citep{kittel2005introduction}. However, in this work, we use them for parameterization of the electron density functions which is the square of the magnitude of the wavefunction.

\subsection{Gaussian-Type Orbital Basis}
Gaussian-type orbitals \citep[GTO]{hehre1972self} offer an alternative to the PW basis for the expression of electronic structures. We consider GTOs in the following form:
\begin{equation}\label{eq:gtobasis}
    \phi_{nlm}(\vx, \vx_{0}) = 
    C_{nlm} R_{nl}(\lVert \vr \rVert)
    Y_{lm}\bigg(\frac{\vr}{\lVert \vr \rVert}\bigg),
\end{equation}
where $\vr = \vx - \vx_{0}$, $\vx_{0}$ is center of the GTO, $C_{nlm}$ is a normalization constant, $R_{nl}(\cdot)$ is the Gaussian radial basis function (RBF), and $Y_{lm}(\cdot)$ is the spherical harmonics (SH). The index of radial basis $n$ and degree $l$ of the spherical harmonics is a non-negative integer, and the magnetic order $m$ is an integer satisfying~$-l \le m \le l$. The GTO basis does not satisfy the periodic boundary condition \cref{eq:pbc}.


The use of Gaussian-type orbitals in quantum chemistry has become popular due to their mathematical simplicity and efficiency in electronic structure computation \citep{hehre1972self}. They allow straightforward integrations and often result in faster convergence in numerical simulations compared to the Slater-type orbitals \citep{gill1994molecular}.




\setlength{\belowcaptionskip}{-0.5em}
\begin{figure*}[t]
    \centering
    \includegraphics[width=0.9\linewidth]{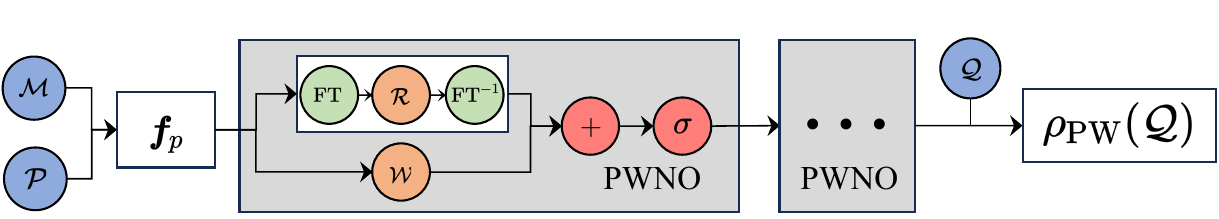}
    \caption{\textbf{Neural architecture for the PWNO layer.} Starting from a molecule $\gM$, our framework makes a PW-based prediction $\rho_{\text{PW}}$ for given query points $\gQ$. The PWNO layer processes the discretized signal using convolution, which is expressed as applying FT, frequency-wise linear transformation $\mathcal{R}$ and inverse FT in sequence. Point-wise activation function $\sigma$ is introduced for the non-linear transformation of the discretized signals.}
    \label{fig:method}
\end{figure*}
\setlength{\belowcaptionskip}{0pt}


\section{Method}
\subsection{Overview}
Here, we provide a high-level description of our Gaussian plane wave neural operator (GPWNO). Our main idea is to introduce the plane-wave basis set for effectively representing the electron density functions. However, since the electron density function consists of high-frequency elements near the atoms, we additionally introduce the GTO-based basis set for representing such elements. We depict the visualization of the overall architecture in \cref{sec:arch} and we provide the complexity analysis of our model in \cref{sec:complex}

\textbf{Decomposition of prediction.} We parameterize our electron density prediction $\rho_{\text{pred}}: \mathbb{R}^{3} \rightarrow \mathbb{R}$ as follows:
\begin{equation}
    \rho_{\text{pred}}(\vx) = \rho_{\text{PW}}(\vx) + \rho_{\text{GTO}}(\vx),
\end{equation}
where $\rho_{\text{PW}}$ and $\rho_{\text{GTO}}$ correspond to our prediction using the plane-wave basis set and the GTO-based basis set, respectively. For training, we use the mean-squared error:
\begin{equation}
    \mathcal{L}_{\mathcal{M}} = \frac{1}{|\mathcal{D}_{\mathcal{M}}|}\sum_{\vx \in \mathcal{D}_{\mathcal{M}}}\lVert \rho(\vx) - \rho_{\text{pred}}(\vx)\rVert^2,
\end{equation}
where we train over different molecules $\mathcal{M}\in\mathcal{D}$ and the locations $\vx\in\mathcal{D}_{\mathcal{M}}$ is the dataset of locations where the electron density is known, e.g., computed by DFT simulations.

\textbf{PBC implementation.} Importantly, our parameterization guarantees the predictions to satisfy the periodic boundary conditions on PW and GTO-based prediction. Specifically, we utilized the crystal graph framework introduced by \citet{Xie18}, which emulates periodic boundary conditions common in computational physics, as outlined in \citet{landau2021guide}. In simple terms, we consider each atom to have multiple coordinates, each translated by integer multiples of lattice vector, while maintaining identical node features during message passing.

\textbf{Equivariance.} Both the electron density function and our predictions are scalar fields that are SE(3)-equivariant. In particular, our PW-based prediction is a scalar field since the PW basis aligns with the lattice, which transforms along with the molecule and hence the PW coefficients are invariant.\footnote{We provide proof for the aperiodic molecule in \cref{sec:eqproof}} The proof for the GTO-based prediction being a scalar field follows that of \citet{cheng2023equivariant}.

\textbf{Masking for decomposition.} Moreover, the density prediction $\rho_{\text{PW}}(\vx)$, as characterized by the PW basis set, is designed to be zero within a specific cutoff radius for certain atoms. This approach aids in segmenting the target region for each basis and integrates the physical understanding that high-frequency electron density components are indicative of core (non-valence) electrons, which are typically located near an atom's center.

\subsection{PW-based Prediction}
Here, we further describe our PW-based prediction $\rho$ of the electron density. To this end, following the Fourier neural operator \citep[FNO]{li2020fourier}, we let each layer of our network implicitly represent a signal updated in both the Fourier space and the Euclidean space. However, unlike the FNO approach, our signal is represented using the PW basis and the corresponding coefficients, which is a subset of Fourier series that satisfies the periodic boundary condition. 

\textbf{Signal initialization.} To be specific, we initialize the first layer using composition of tensor field network \citep[TFN]{thomas2018tensor}, grid point-wise embedding, and discrete Fourier transformation. We first obtain the atom-wise embedding of the molecule as $\mathbf{f}_{u}^{(0)} = \operatorname{TFN}_{u}(\mathcal{M})$, where $\operatorname{TFN}_{u}$ denotes the TFN output associated with the atom~$u\in\mathcal{V}$. We provide more details of our TFN architecture in \cref{sec:siginit}. 

\textbf{Lattice-based discretization.} For point-wise embedding, we locate probe nodes $\mathcal{P}$ on a three-dimensional space inside the lattice $\mathbf{L} = [\va_{1}, \va_{2}, \va_{3}]^{\top}$. Specifically, we uniformly discretized the space inside the lattice and located the probe node $p$ at each point. i.e., a probe node $p \in \mathcal{P}$ is placed at 
\begin{equation}
    \vx_p=\frac{d_1}{D}\va_1 + \frac{d_2}{D} \va_2 + \frac{d_3}{D}\va_3,
\end{equation} 
where $d_{1}, d_{2}, d_{3} < D$ are non-negative integers and the total number of probe nodes is $M=D^3$. Note that this lattice generalizes the regular voxel grids with cubic cells.

Then, we aggregate messages from the atoms $u\in \mathcal{V}$ in vicinity of a probe point $p\in \mathcal{P}$:
\begin{equation}\label{eq:probe1}
   \mathbf{f}_{p}^{(1)} = \operatorname{UPD} \circ \operatornamewithlimits{AGG} (\{(r_{up}, \mathbf{f}_{u}^{(0)}): u \in{\mathcal{N}_{\mathcal{V}}}(p)\}),
\end{equation}
where $r_{up}= \|\vx_{p}-\vx_{u}\|$ and $\mathcal{N}_{\mathcal{V}}(p)\subseteq\mathcal{V}$ is the set of atoms within a certain radius for the probe node $p\in\mathcal{P}$. Furthermore, $\operatorname{AGG}$ and $\operatorname{UPD}$ are the aggregation and the update function typically used for graph neural networks. The grid embeddings $\{\mathbf{f}_{p}^{(1)}: p \in \mathcal{P}\}$ implicitly represent a signal $f^{(1)}$ evaluated at location $\vx_{p}$, i.e., $\mathbf{f}_{p}^{(1)} = f^{(1)}(\vx_{p})$.

\textbf{PW neural operator layer.} To represent the interaction between the PW bases, we employ the PW neural operator layer (PWNO). PWNO iteratively updates the signal $f^{(h)}:\R^3 \to \R^C$ by a convolution operator using the PW basis where $C$ is the channel dimension of signal, i.e., sequentially applying Fourier transformation, frequency-wise linear transformation, and inverse Fourier transformation:
\begin{equation}
    f^{(h+1)} = \sigma(\mathcal{W} \cdot f^{(h)} + \mathcal{K} \ast f^{(h)}),
\end{equation}
where $\ast$ is the convolution operator, $\mathcal{K}$ is the kernel, $\mathcal{W}$ is the point-wise linear operator, and $\sigma$ is the point-wise non-linearities. i.e., $\operatorname{GELU}$~\citep{hendrycks2016gaussian}. We implement the convolution operator via linear transformation in the Fourier space:
\begin{equation}
    \mathcal{K} \ast f^{(h)} = \operatorname{FT}^{-1}(\mathcal{R} \cdot \operatorname{FT}(f^{(h)})),
\end{equation}
where $\mathcal{R}$ is the frequency-wise linear transformation, $\operatorname{FT}$ is the Fourier transformation that maps point-wise representation into the PW basis coefficients, and $\operatorname{FT}^{-1}$ its inverse. 

To implement the Fourier transformation in discrete space, we operate with point-wise representations $\{\mathbf{f}_{p}^{(h)} \in \mathbb{R}^{C}: p \in \mathcal{P}\}$ and frequency-wise representations $\{\hat{\mathbf{f}}_{\bm{\lambda}}^{(h)} \in \mathbb{R}^{C}: \bm{\lambda} \in \Lambda\}$ of the underlying signal $f^{(h)}$, e.g., 
$f^{(h)}(\bm{x}) = \sum_{\bm{\lambda}\in\Lambda}\hat{\mathbf{f}}_{\bm{\lambda}}^{(h)} \phi_{\bm{\lambda}}(\bm{x})$
where $\phi_{\bm{\lambda}}(\bm{x})$ is the PW with wavevector $\bm{\lambda}$.
Our framework operates only on the wavevectors $\Lambda$ of the PW basis, incorporating the periodicity constraints regarding the lattice structure $\Lambda$. We depicted the architecture of the PWNO layer in \cref{fig:method}.

\textbf{Prediction network.} At the final $H$-th layer, given a query point $\bm{x}_q$, we compute the output as follows:
\begin{equation}
    \label{eq:query1}
    \mathbf{f}_{q}^{(H)} = \operatorname{UPD}\circ \operatornamewithlimits{AGG} (\{(r_{pq},\mathbf{f}_{p}^{(H-1)}): p \in \gN_{\mathcal{P}}(q)\}),
\end{equation}
where $r_{pq}=\|\vx_q - \vx_p\|$ and $\gN_{\mathcal{P}}(q) \subseteq{\mathcal{P}}$ is the set of probe nodes within a certain radius for the query node $q$. Note that the query node $p$ can be placed at any points in the continuous space unlike the probe node $p$ which is regularly placed along the lattice. One can also process multiple queries in parallel without re-computing the final probe-wise features $\mathbf{f}_{p}^{(H-1)}$ for different queries.

\begin{table*}[t]
\caption{\textbf{Evaluation of GPWNO for aperiodic materials.} We report the performance for QM9 and MD dataset in NMAE (\%). The best number is highlighted in \textbf{bold}. The baseline results are from \citet{cheng2023equivariant}. Each number is averaged over three runs. For brevity, we denote DimeNet and DeepDFT by DmNet and DDFT, respectively.}
    \centering
    \resizebox{1.0\textwidth}{!}{%
    \begin{tabular}{lccccccccccc}
        \toprule
        Dataset & LNO & FNO & GNO & DmNet++ & DmNet & EGNN & DDFT2 & DDFT & CNN & InfGCN & \textbf{GPWNO} \\
        \midrule
        QM9 & 26.14 & 28.83 & 40.86 & 11.69 & 11.97 & 11.92 & \phantom{0}1.03 & 2.95 & \phantom{0}2.01 & \phantom{0}0.93 & \textbf{0.73} \\
        \midrule
        Ethanol & 43.17 & 31.98 & 82.35 & 14.24 & 13.99 & 13.90 & \phantom{0}8.83 & 7.34 & 13.97 & \phantom{0}8.43 & \textbf{4.00} \\
        Benzene & 38.82 & 20.05 & 82.46 & 14.34 & 14.48 & 13.49 & \phantom{0}5.49 & 6.61 & 11.98 & \phantom{0}5.11 & \textbf{2.45} \\
        Phenol & 60.70 & 42.98 & 66.69 & 12.99 & 12.93 & 13.59 & \phantom{0}7.00 & 9.09 & 11.52 & \phantom{0}5.51 & \textbf{2.68} \\
        Resorcinol & 35.07 & 26.06 & 58.75 & 12.01 & 12.04 & 12.61 & \phantom{0}6.95 & 8.18 & 11.07 & \phantom{0}5.95 & \textbf{2.73} \\
        Ethane & 77.14 & 26.31 & 71.12 & 12.95 & 13.11 & 15.17 & \phantom{0}6.36 & 8.31 & 14.72 & \phantom{0}7.01 & \textbf{3.67} \\
        MDA & 47.22 & 34.58 & 84.52 & 16.79 & 18.71 & 12.37 & 10.68 & 9.31 & 18.52 & 10.34 & \textbf{5.32} \\
        \bottomrule
    \end{tabular}%
    }
    \label{tab:aperiodic}
\end{table*}

\newcommand{\addpic}{\includegraphics[width=6em]{example-image}}
\newcommand{\addpicv}[1]{\includegraphics[trim={1.35cm 1.35cm 1.35cm 1.35cm}, clip, width=4.6em]{#1}}
\newcommand{\addpicqm}[1]{\includegraphics[trim={2.7cm 2.7cm 2.7cm, 2.7cm}, clip, width=4.6em]{#1}}
\newcommand{\addpich}[1]{\includegraphics[trim={1.4cm 1.4cm 1.4cm 1.4cm}, clip, width=4.6em]{#1}}
\newcommand{\addpicp}[2]{\includegraphics[trim={{#2}cm {#2}cm {#2}cm {#2}cm}, clip, width=4.6em]{#1}}
\newcolumntype{C}{>{\centering\arraybackslash}m{4.6em}}

\begin{figure*}[t]
\centering
\setlength\tabcolsep{1pt} 
\resizebox{\textwidth}{!}{
\begin{tabular}{l*9{C}@{}}
\toprule
 & \multicolumn{3}{c}{QM9 (\ce{O2C3H4})} & \multicolumn{3}{c}{Monoclinic (\ce{Sr(P3Pt2)2})} & \multicolumn{3}{c}{Tetragonal(\ce{Eu(ZnGe)2})} \\
 \cmidrule(lr){2-4} \cmidrule(lr){5-7} \cmidrule(lr){8-10}& GT & \textbf{GPWNO} & InfGCN & GT & \textbf{GPWNO} & InfGCN & GT & \textbf{GPWNO} & InfGCN \\ 
\midrule
\multirow{1}*{Pred.} 
& \addpicqm{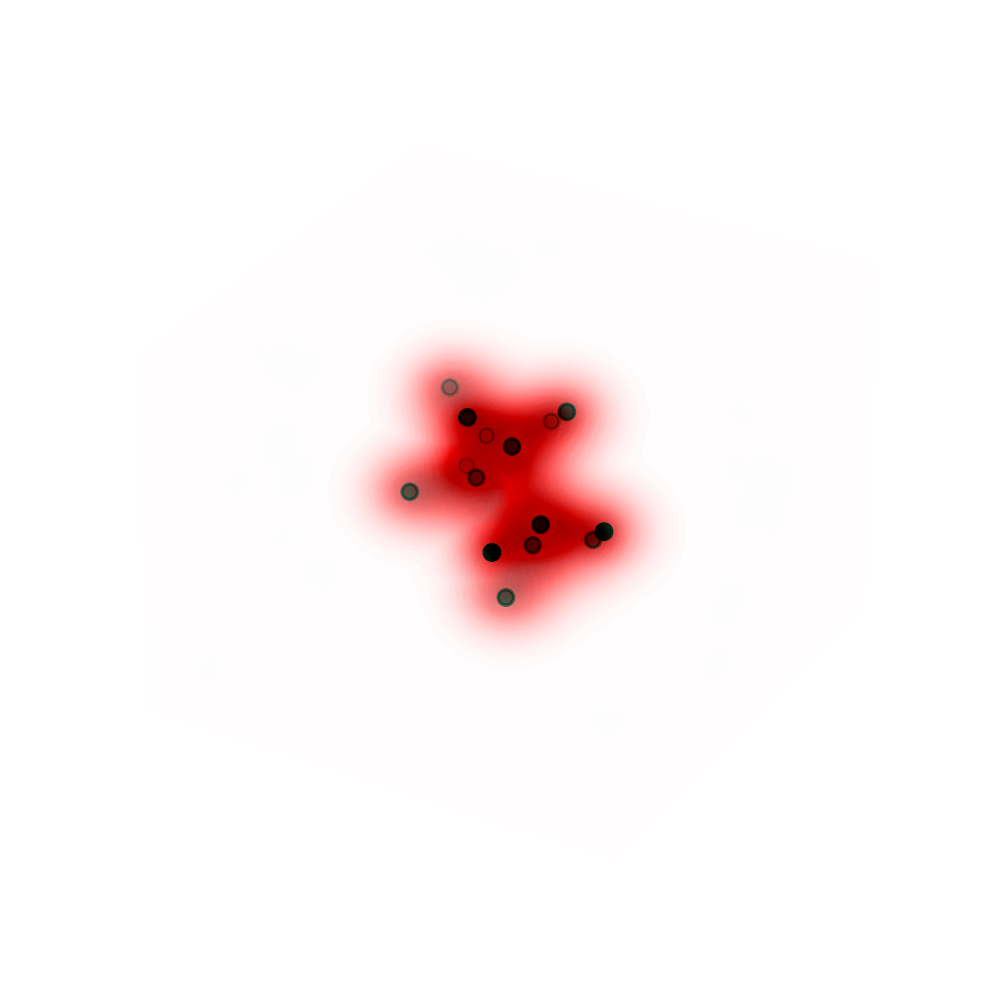}  
& \addpicqm{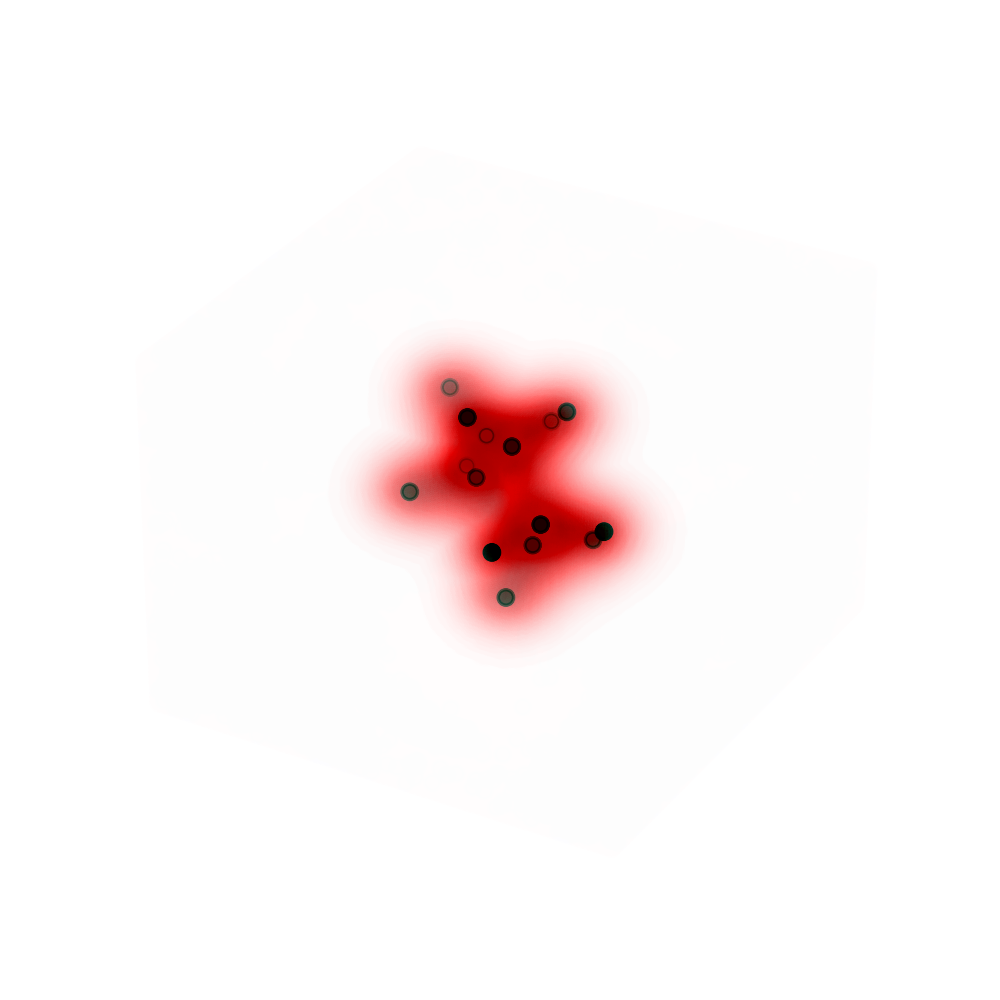}  
& \addpicqm{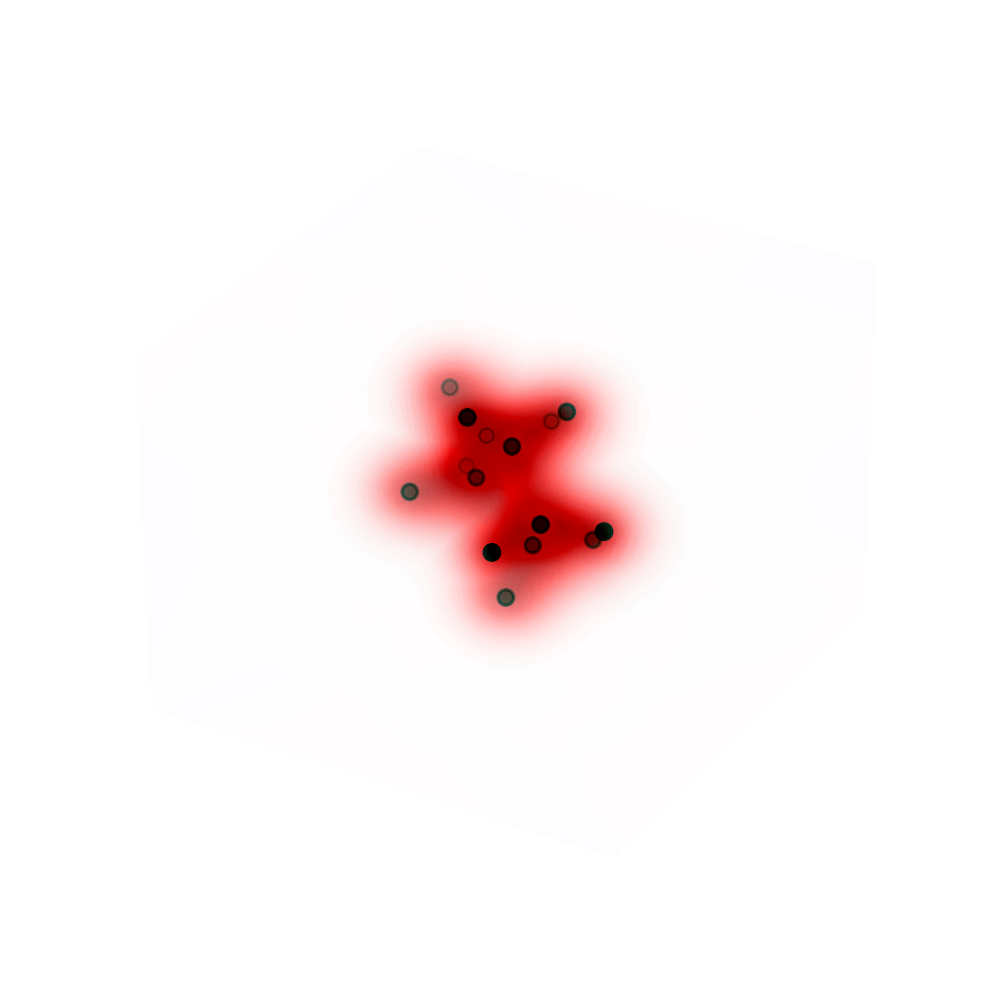} 
& \addpicv{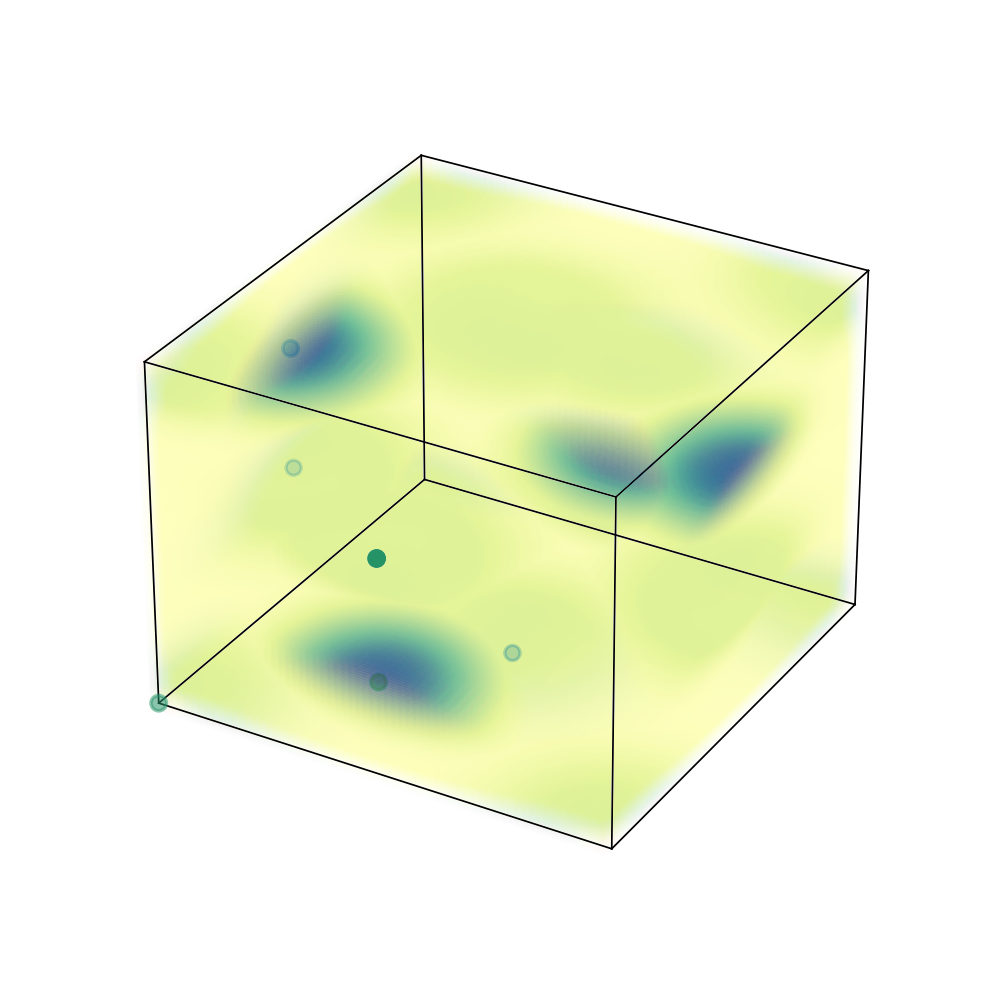}  
& \addpicv{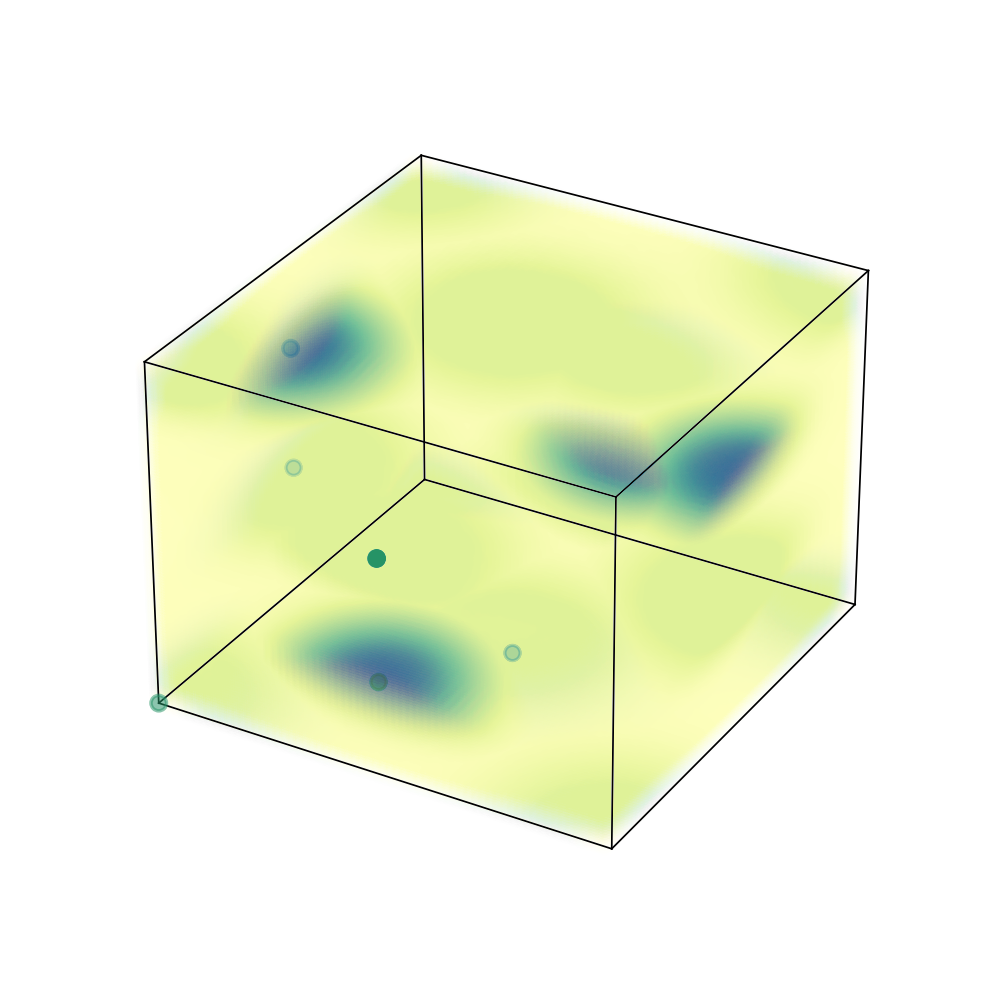}  
& \addpicv{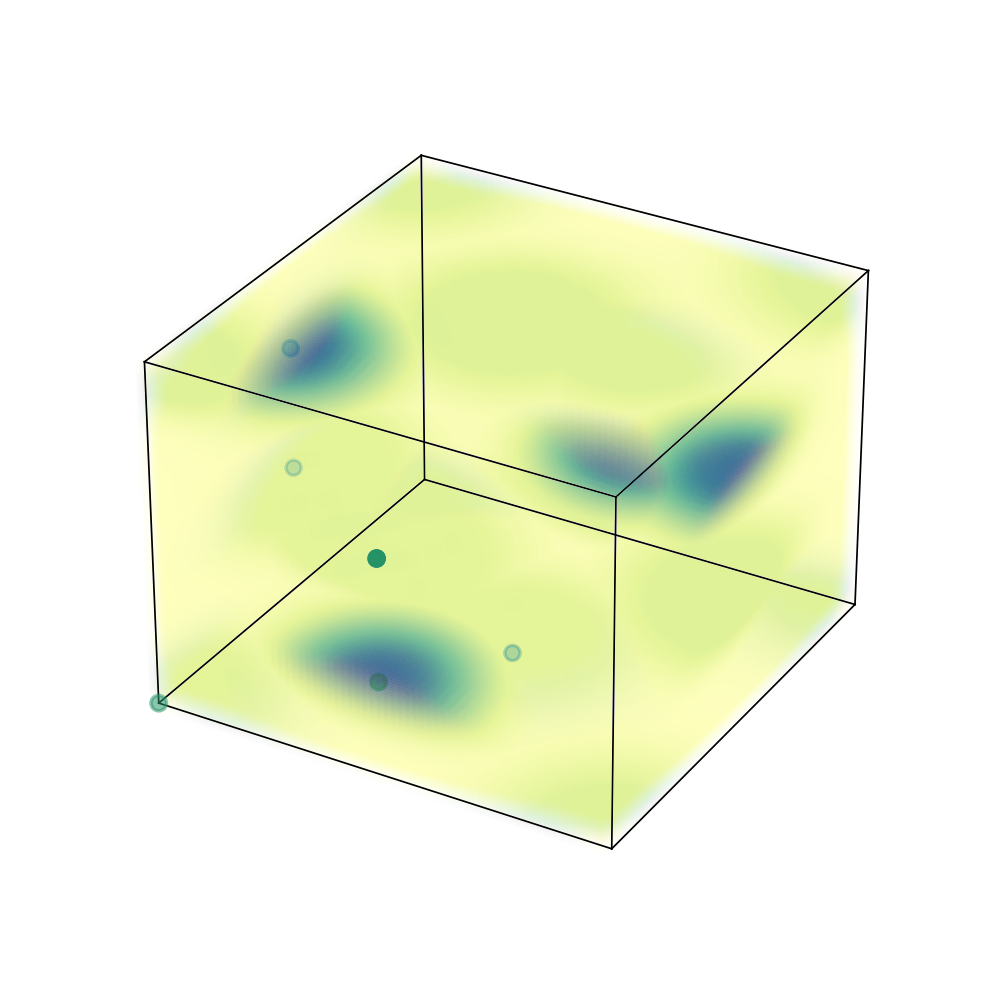}
& \addpich{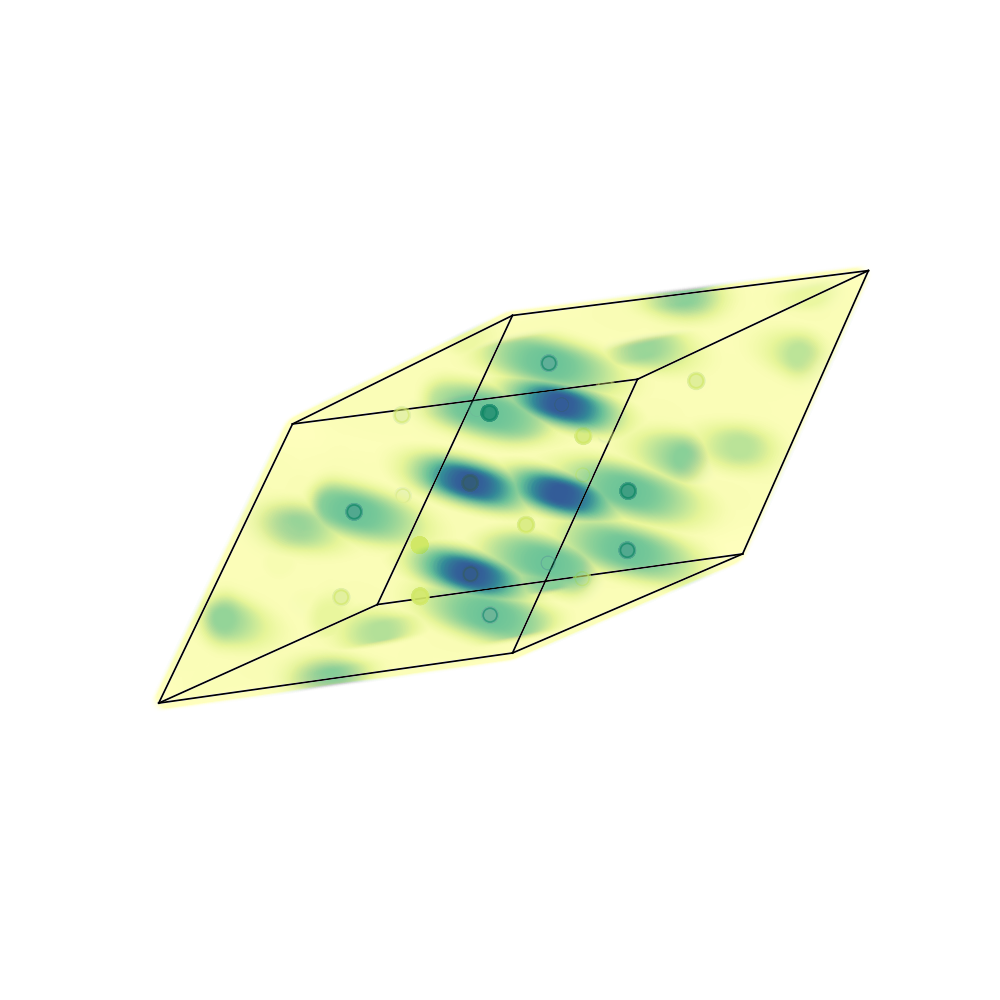}  
& \addpich{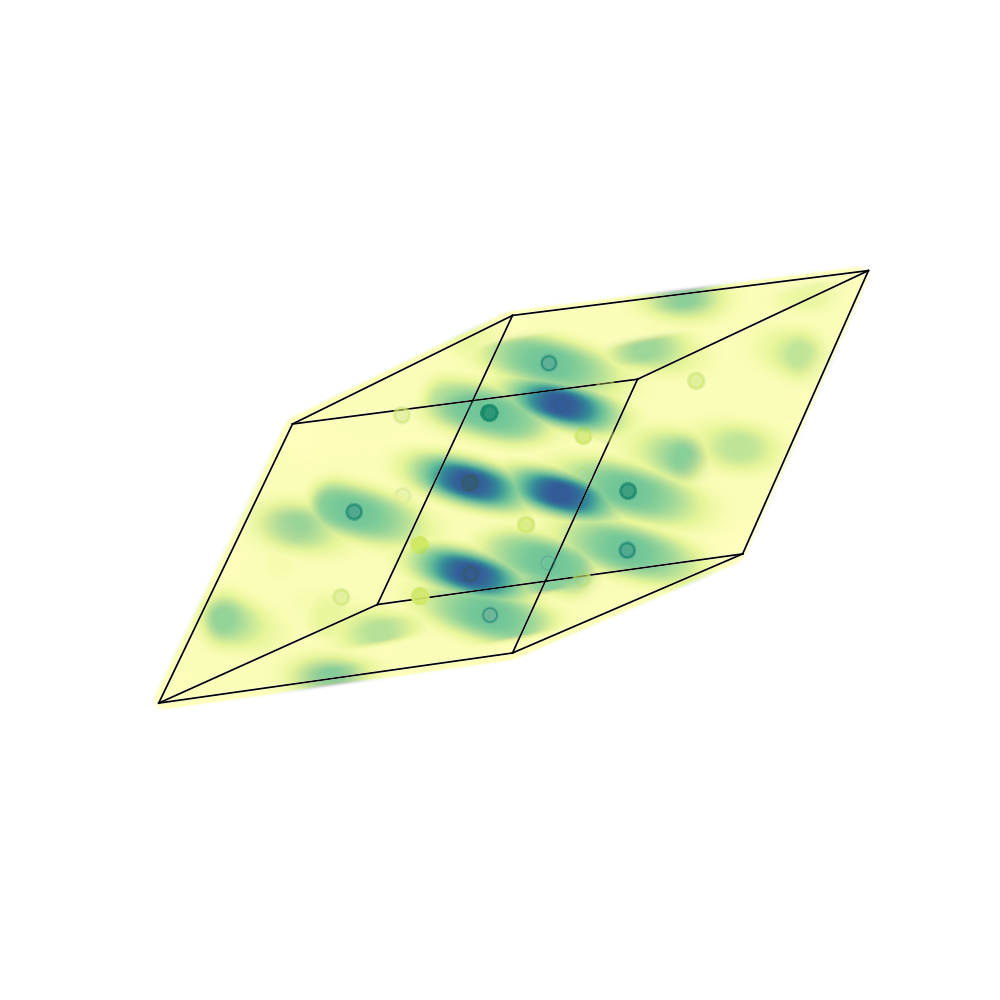}  
& \addpich{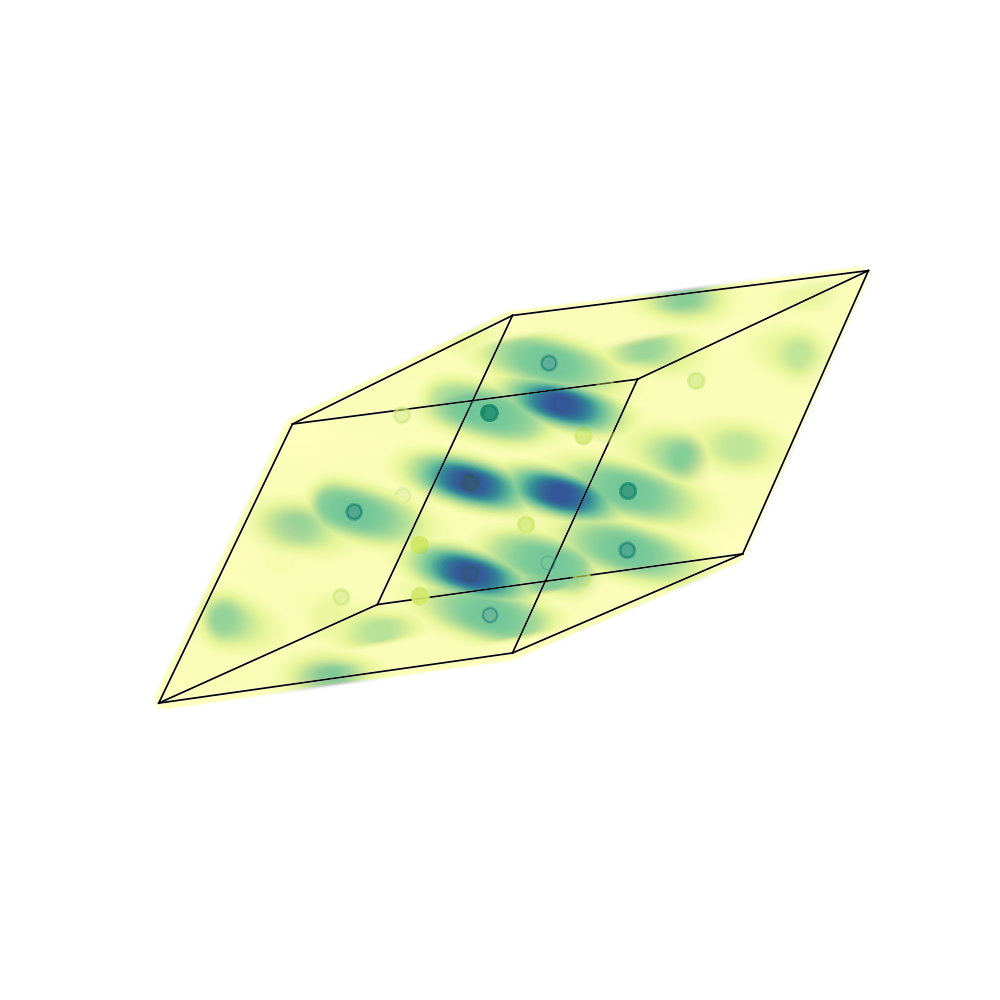} \\ 
Error 
& ~ 
& \addpicqm{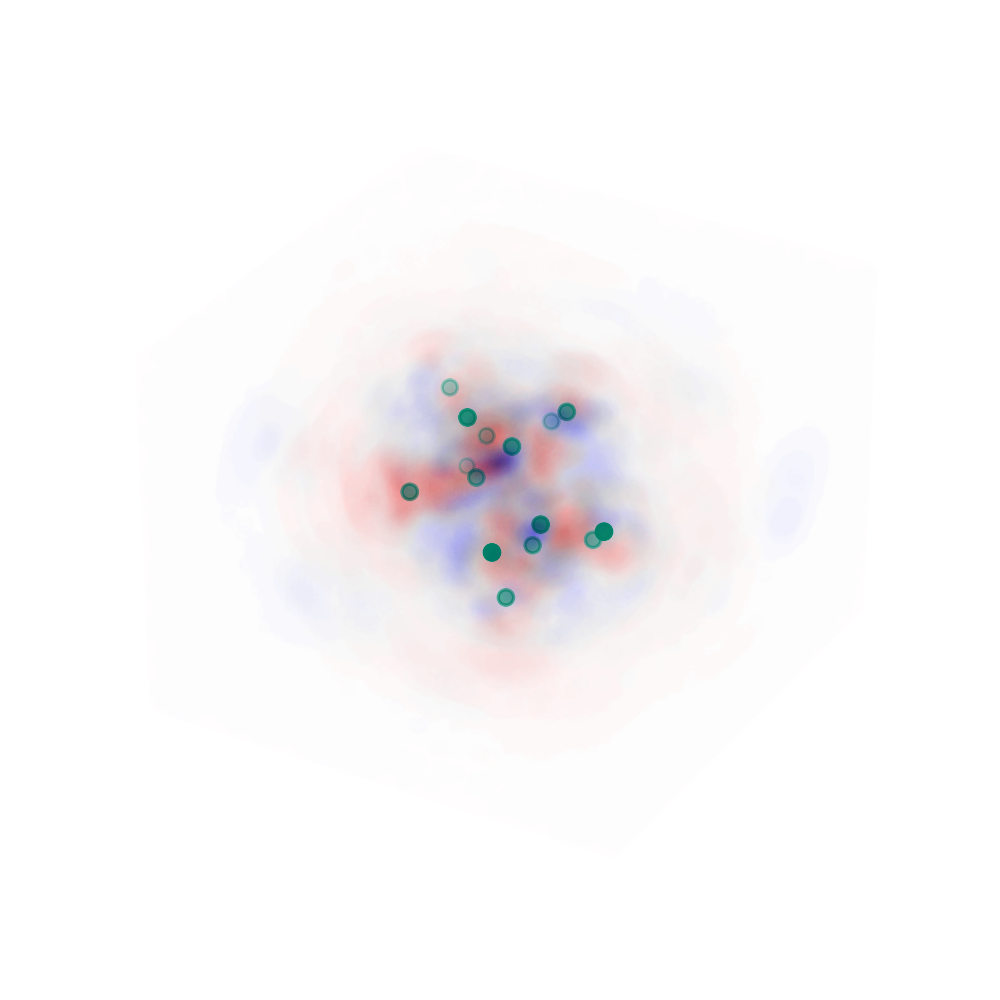}
& \addpicqm{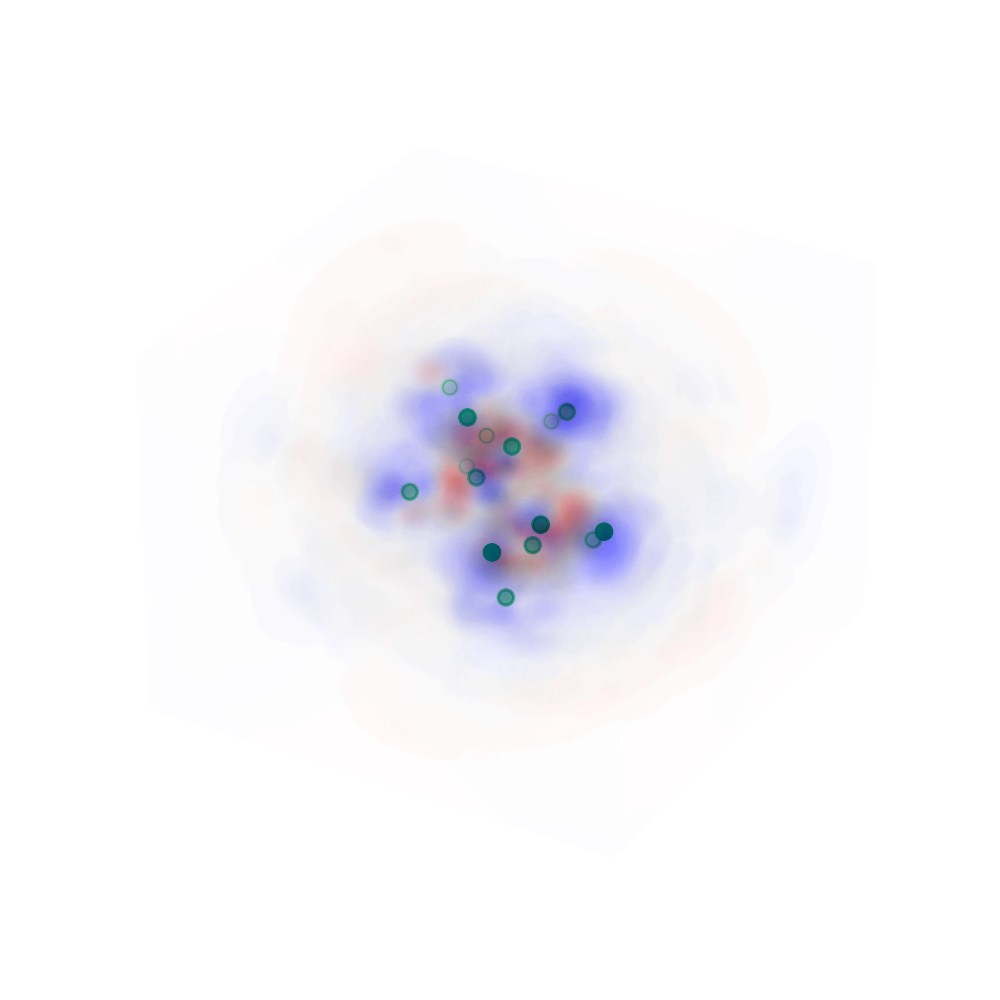}
& ~ 
& \addpicv{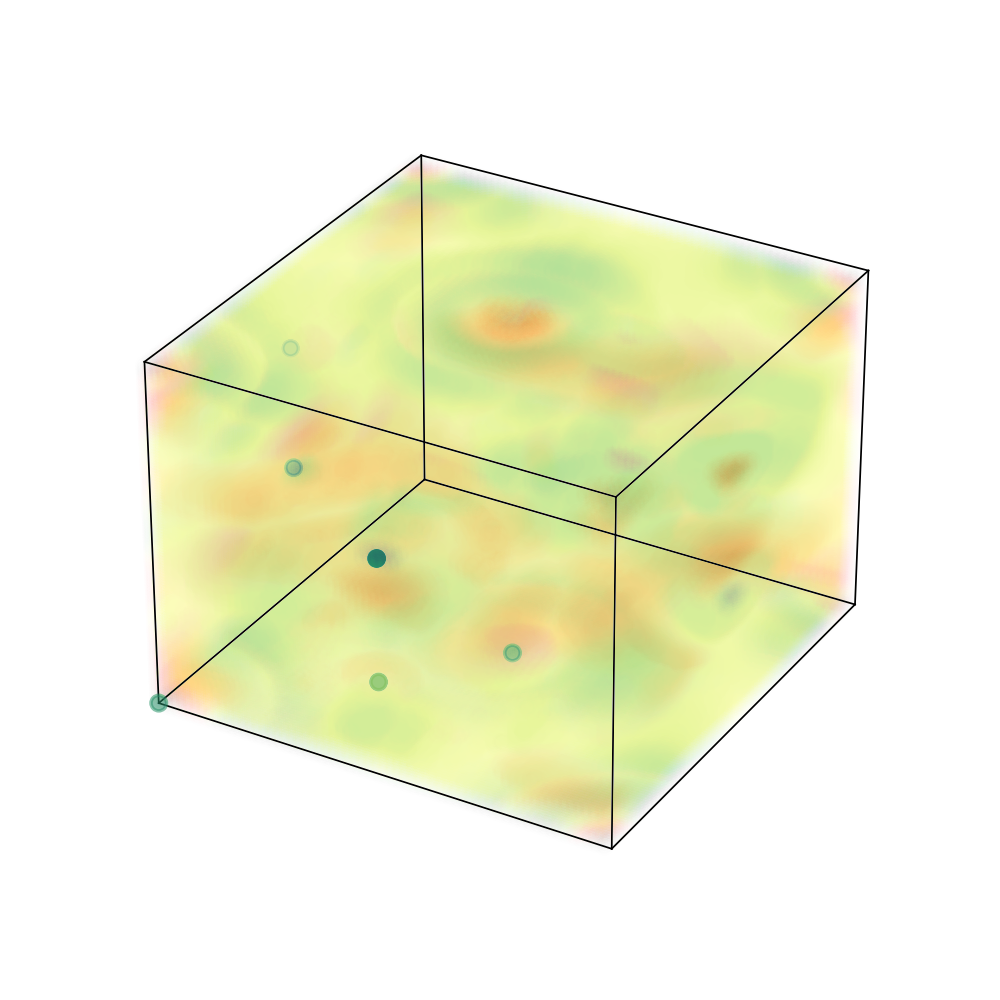}
& \addpicv{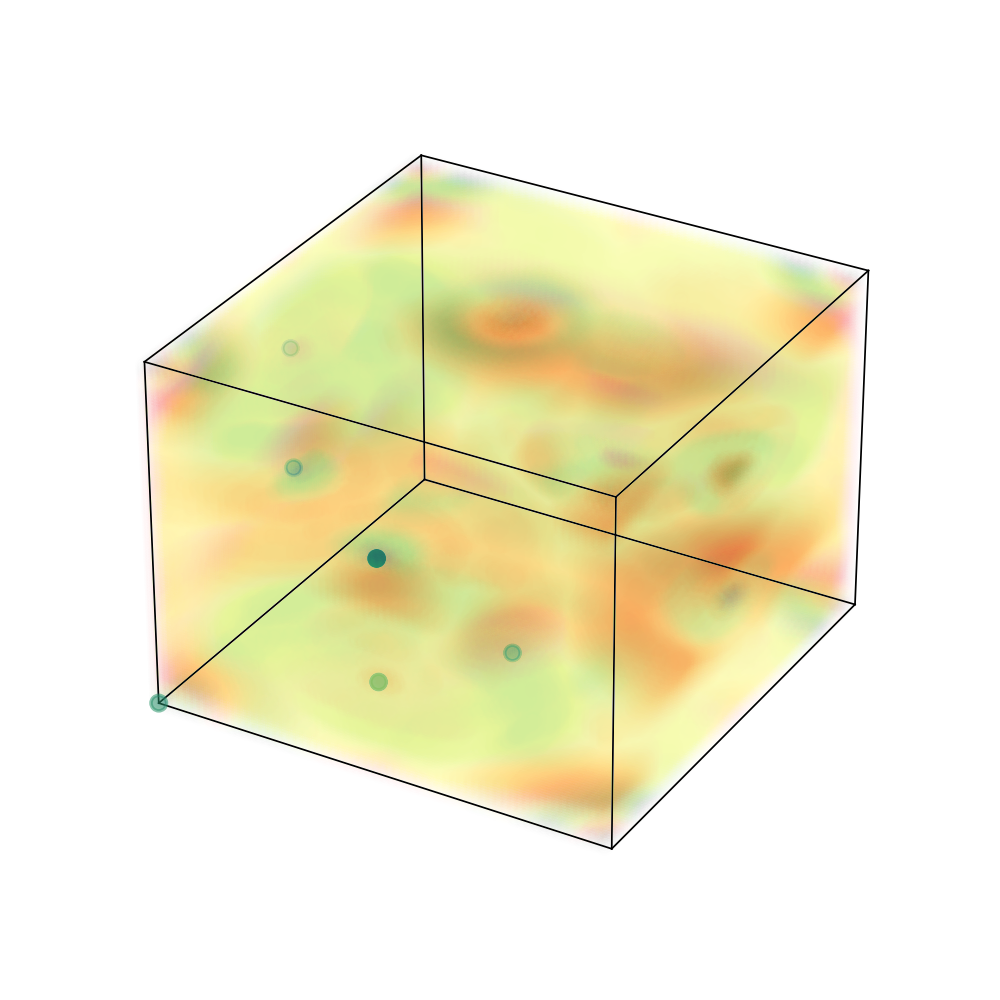}
& ~ 
& \addpich{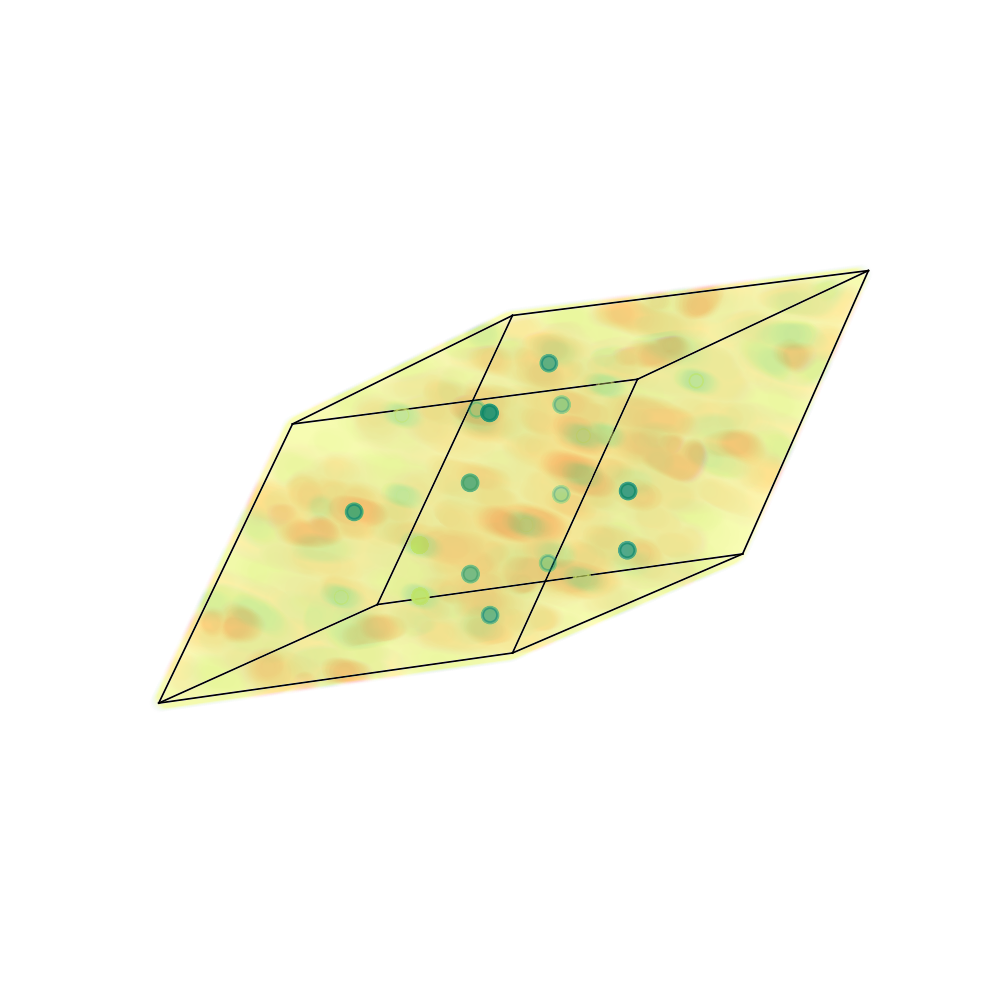}
& \addpich{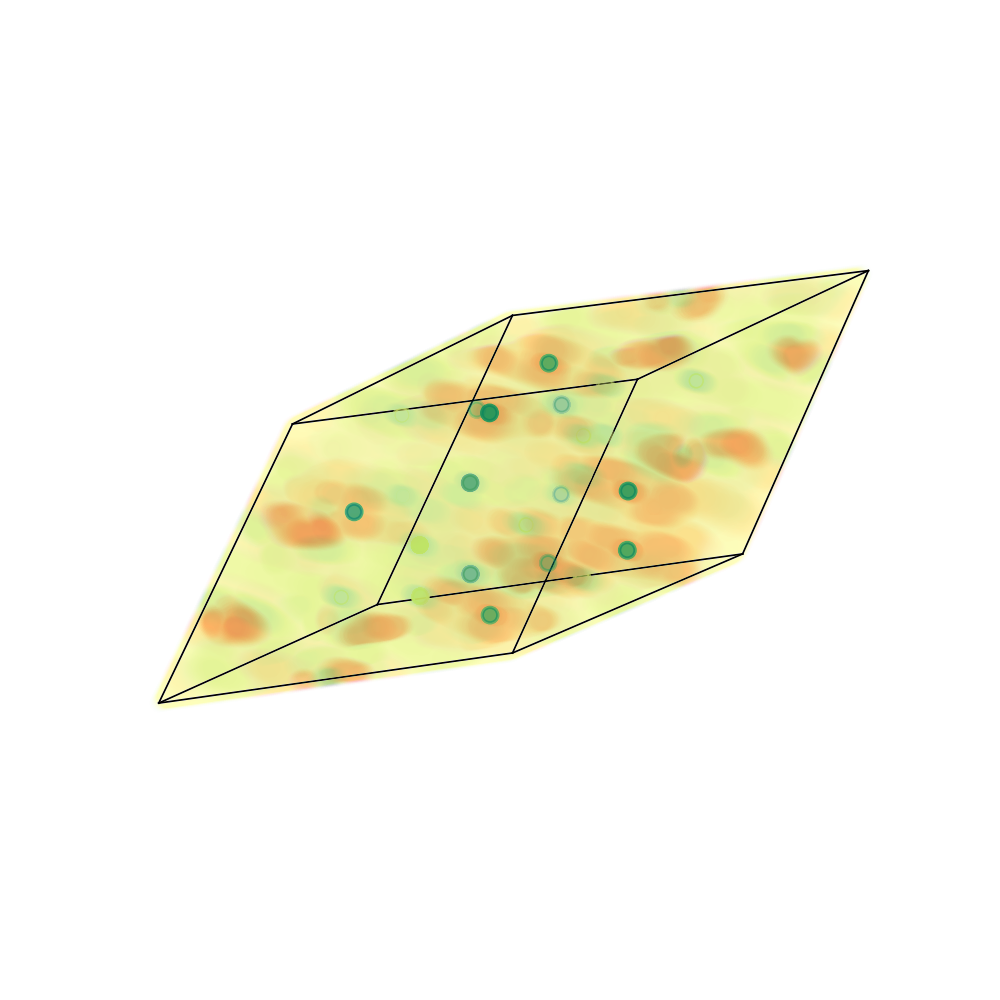}\\ 
\midrule
NMAE &~&\textbf{0.47}&0.59&~&\textbf{5.48}&5.78 &~&\textbf{4.97}&5.28\\
\bottomrule 
\end{tabular}
}\\
\vspace{.05in}
\resizebox{\textwidth}{!}{
\begin{tabular}{l*9{C}@{}}
\toprule
 & \multicolumn{3}{c}{QM9 (\ce{C3ONH5})} & \multicolumn{3}{c}{Triclinic (\ce{Li9Mn2Co5O16})} & \multicolumn{3}{c}{Orthorhombic (\ce{Dy5Sn11})} \\
\midrule
\multirow{1}*{Pred.} 
& \addpicqm{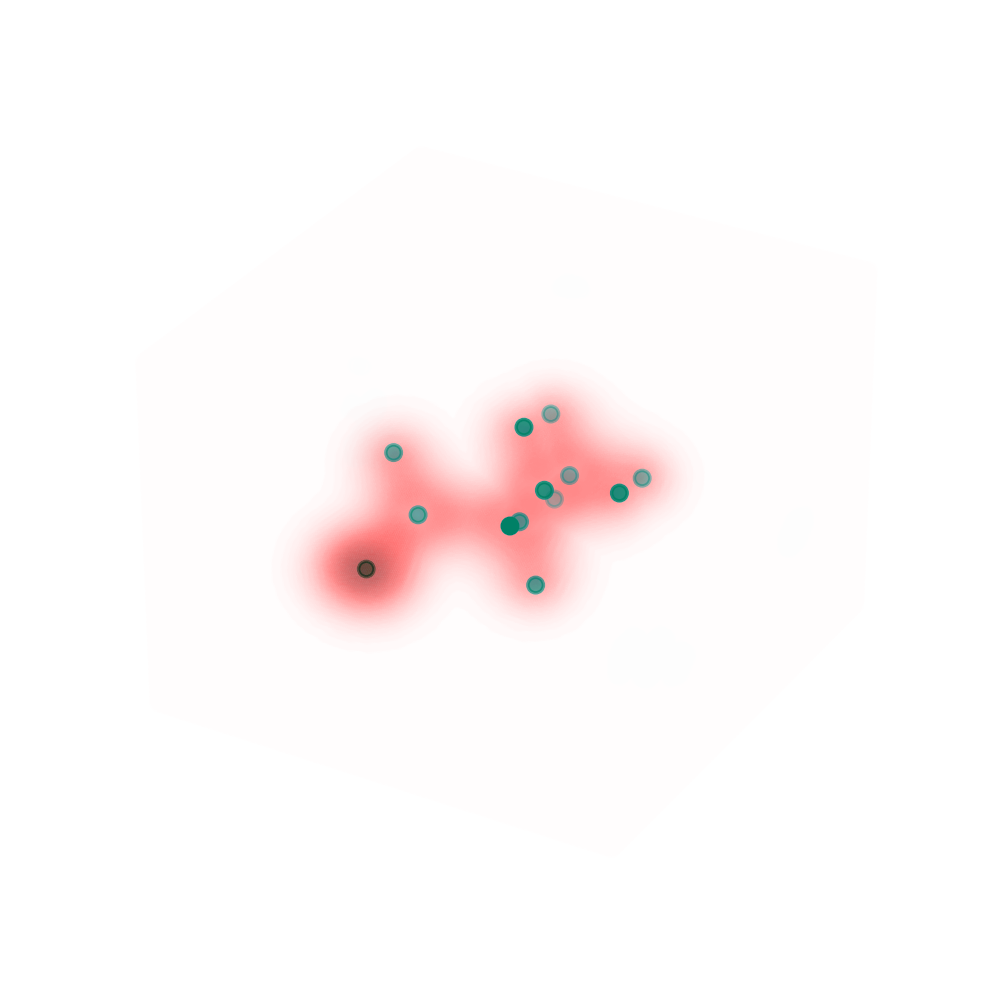}  
& \addpicqm{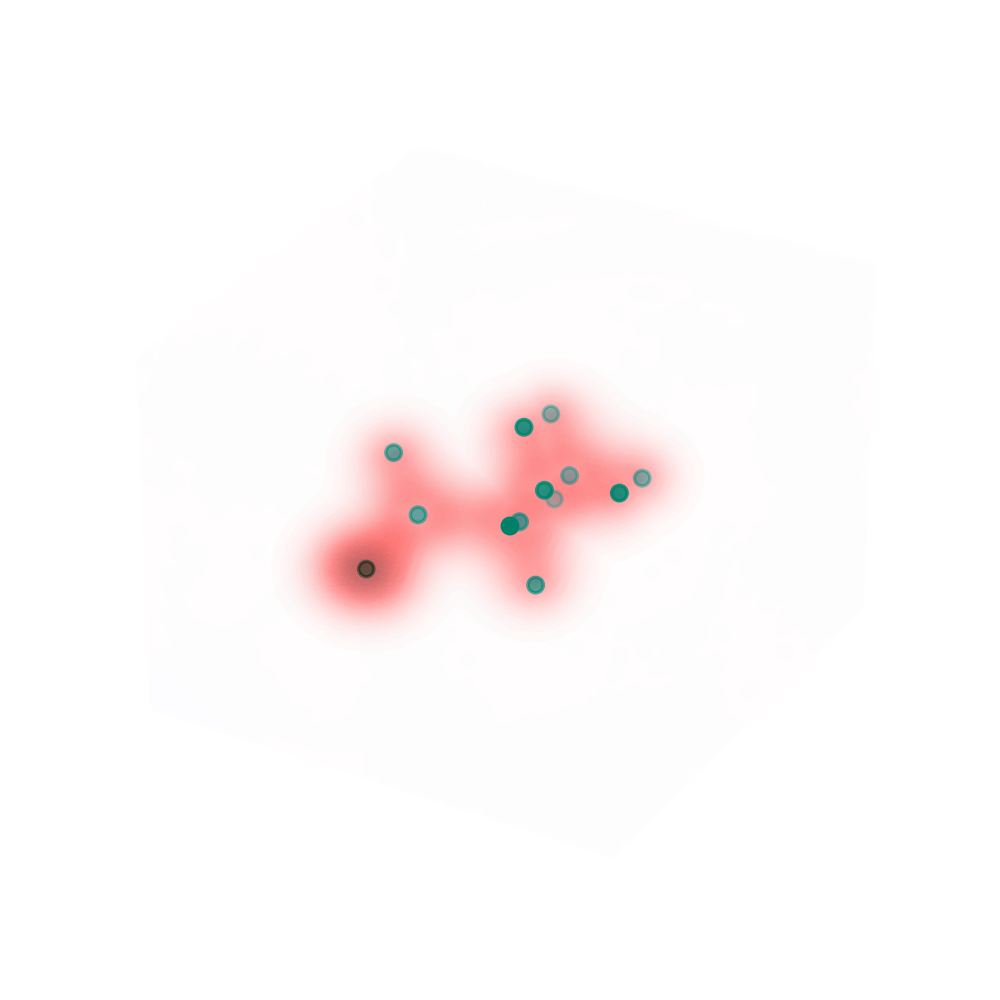}  
& \addpicqm{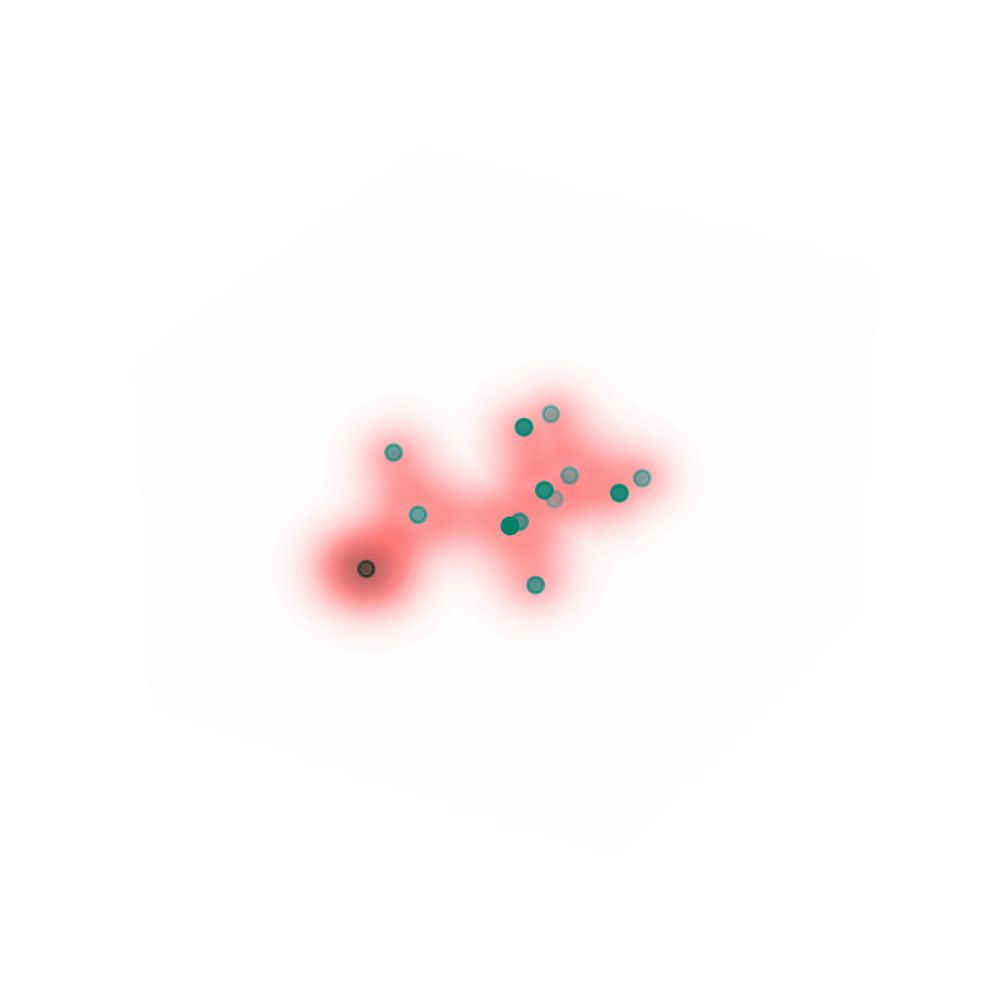} 
& \addpicv{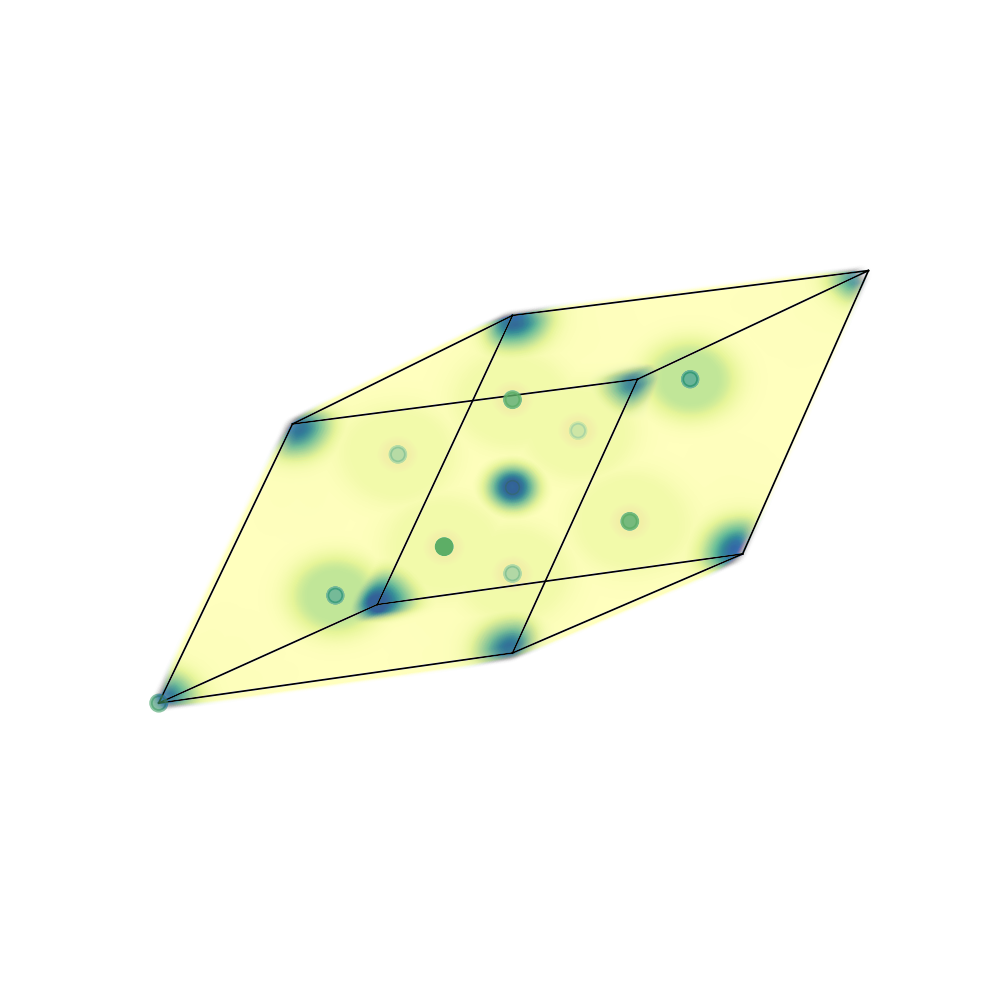}  
& \addpicv{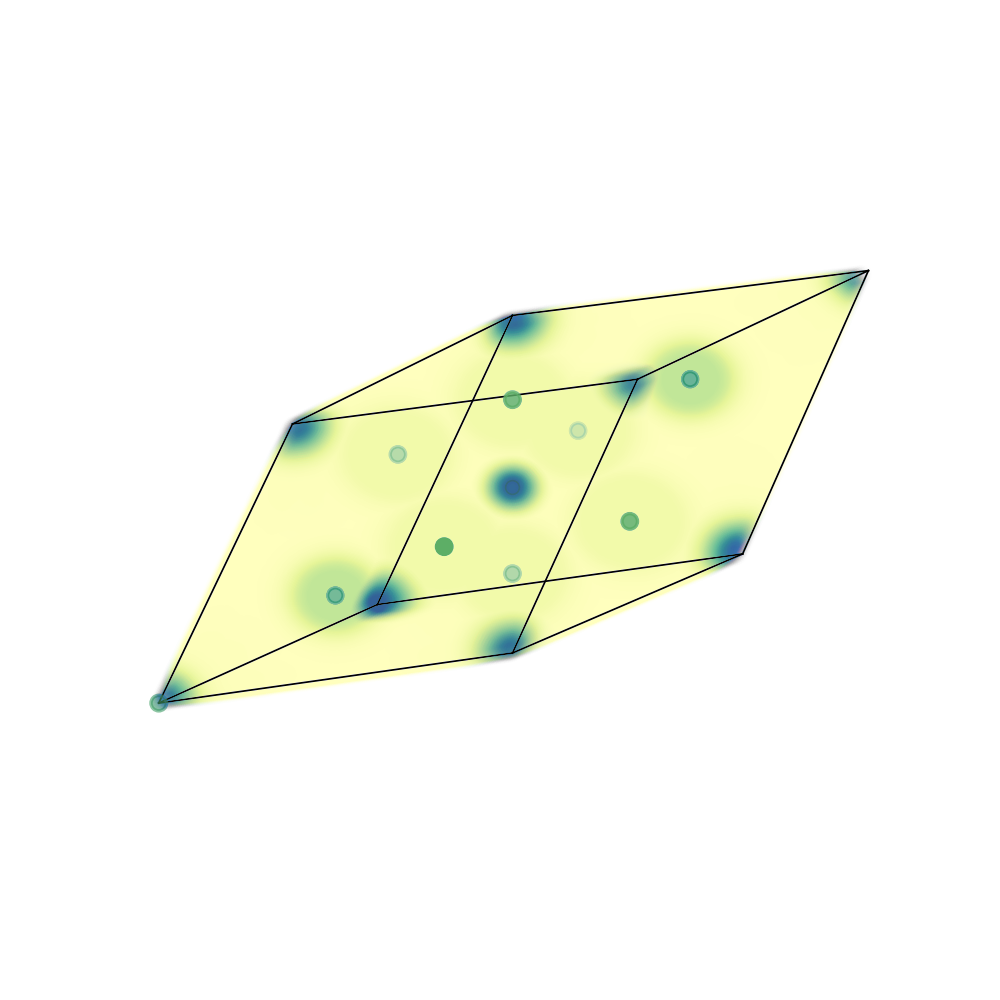}  
& \addpicv{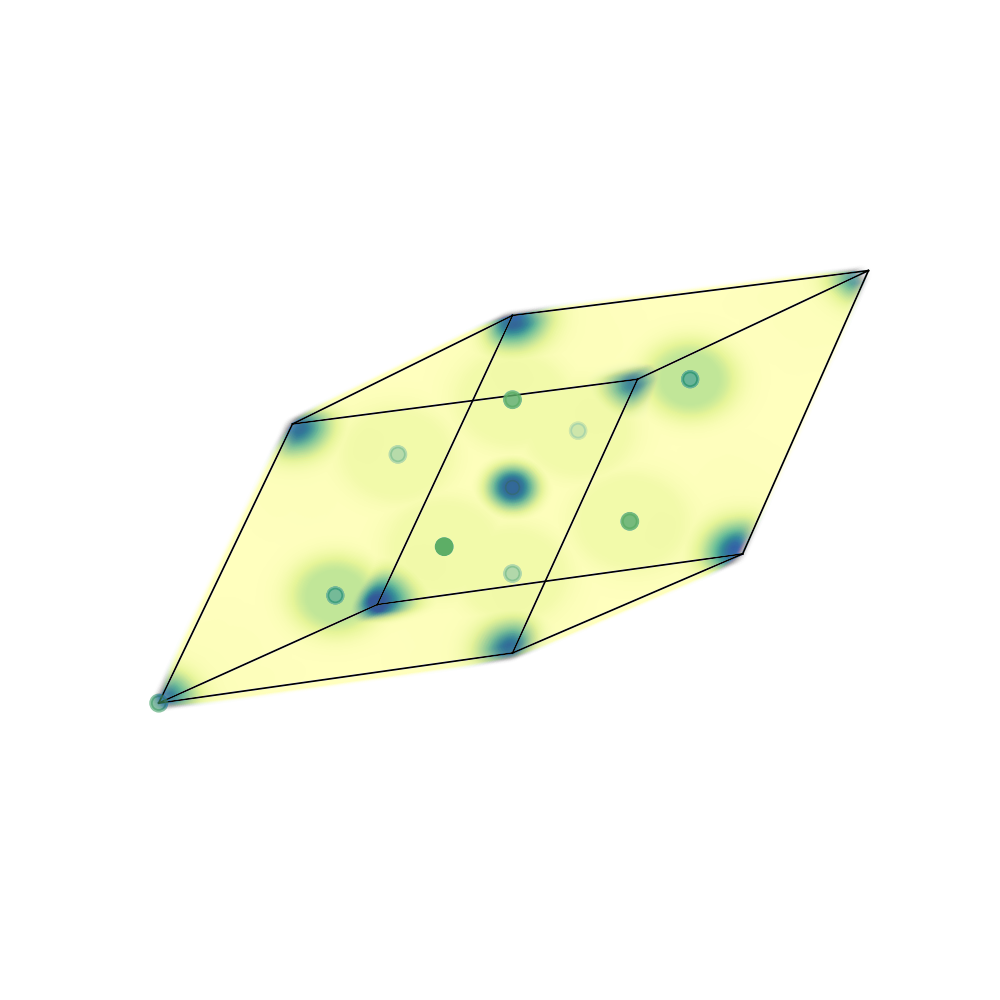}
& \addpich{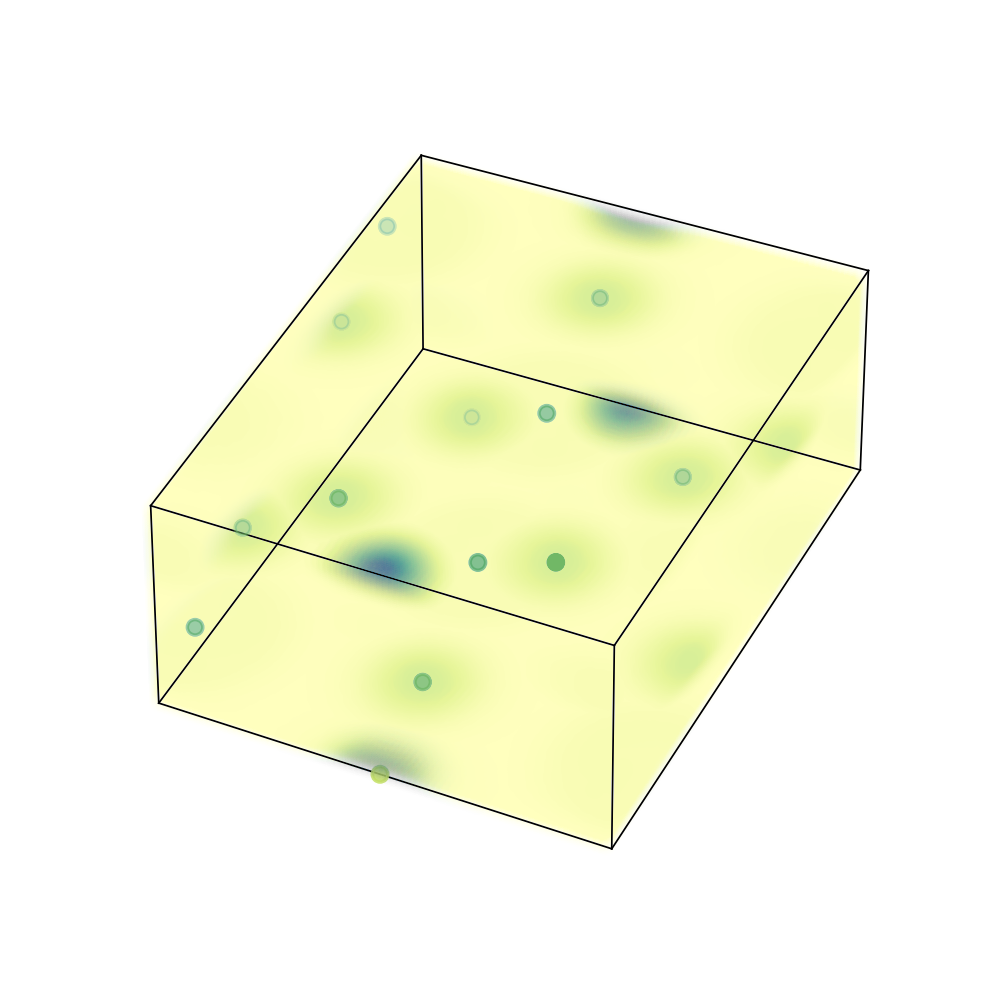}  
& \addpich{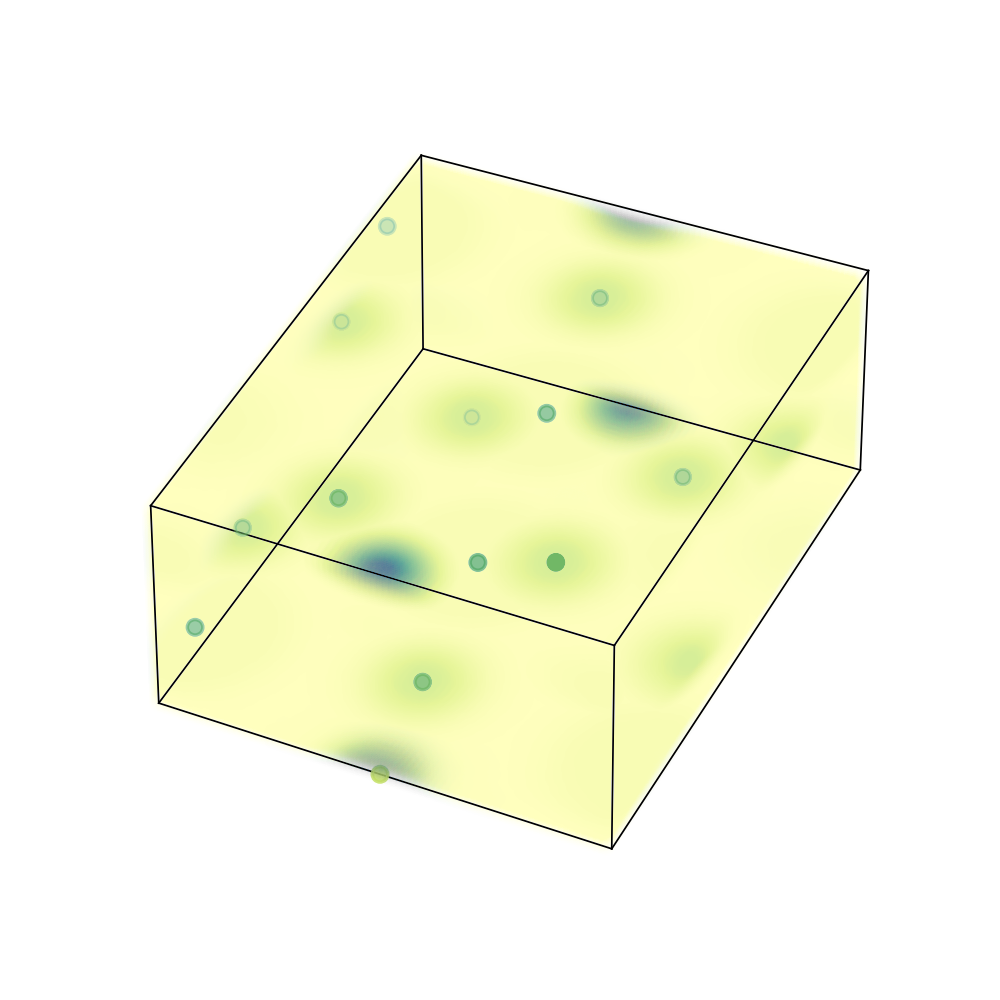}  
& \addpich{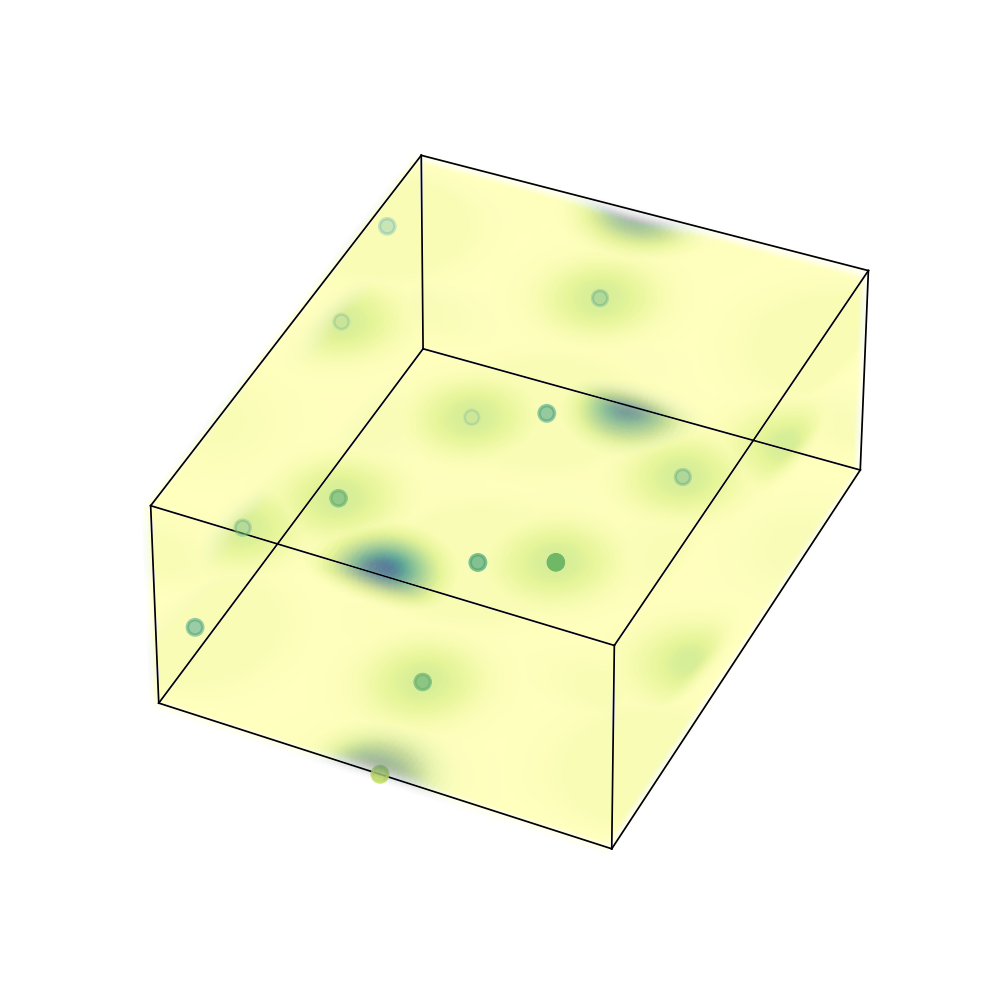} \\
Error 
& ~ 
& \addpicqm{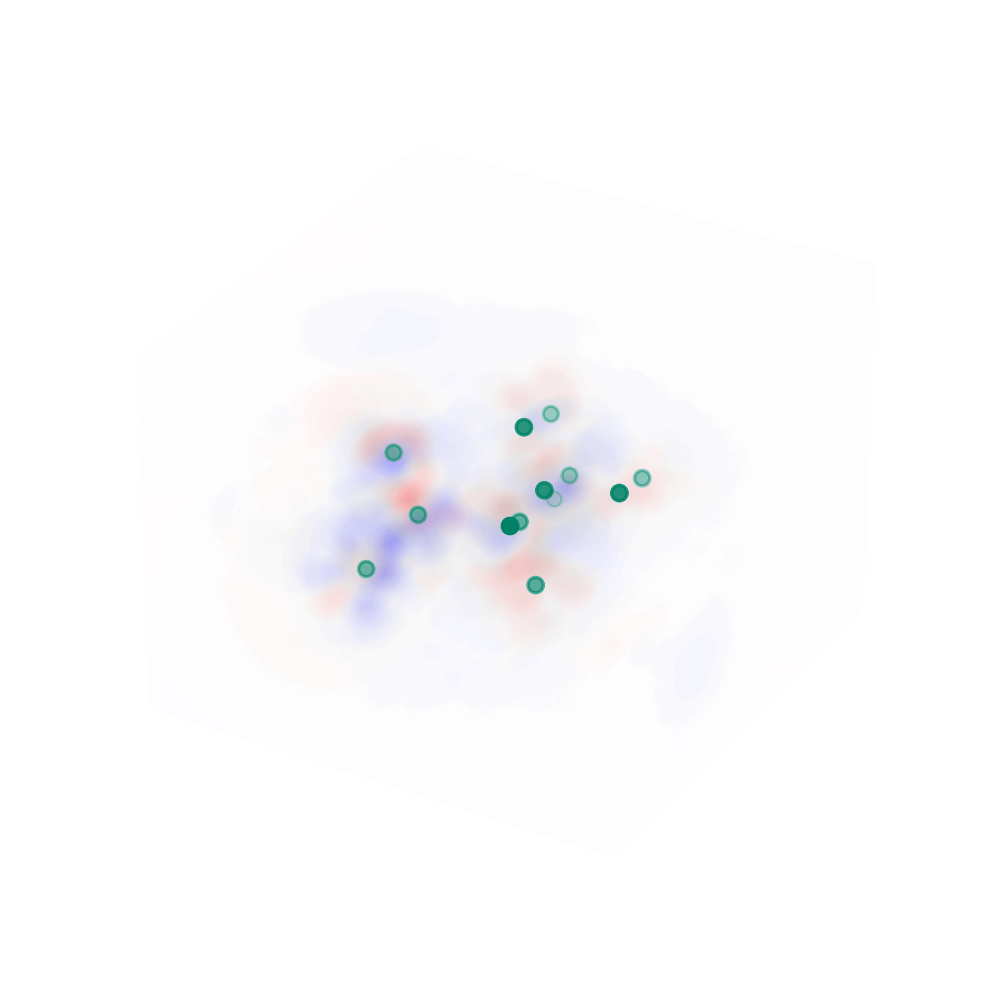}
& \addpicqm{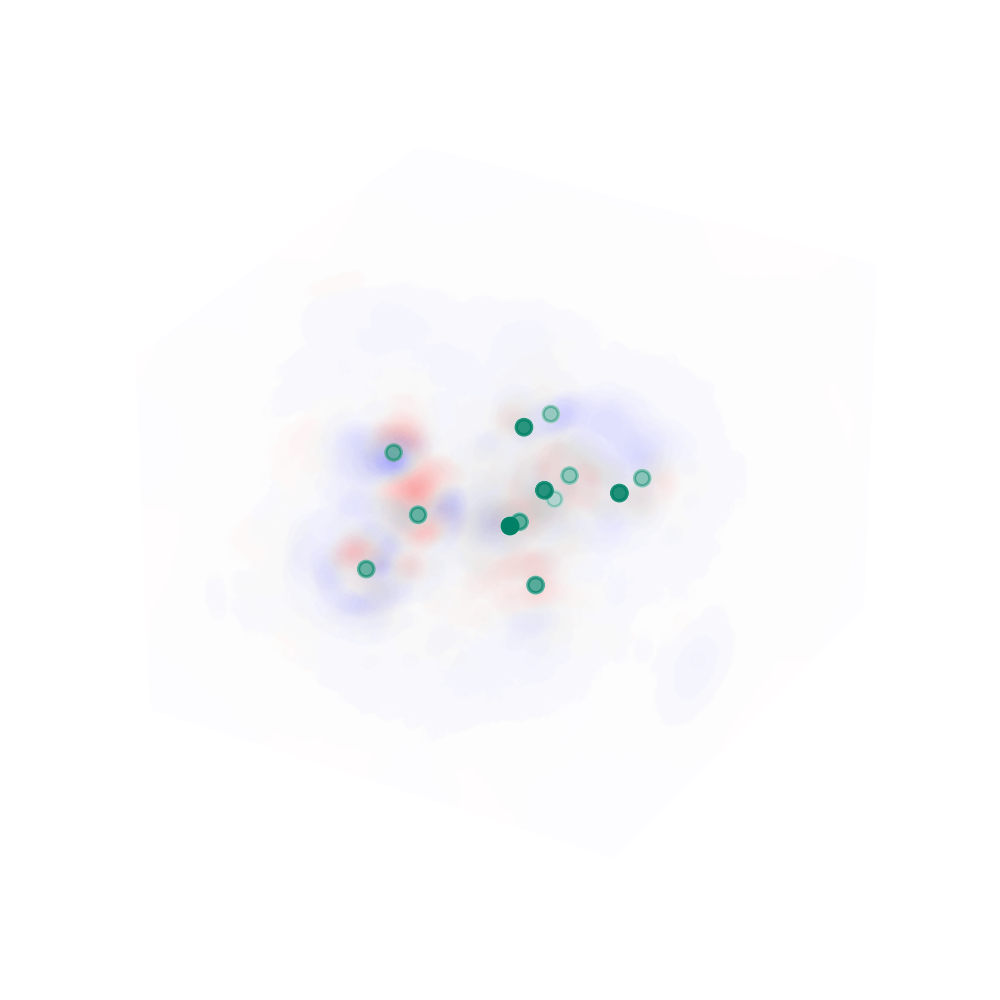}
& ~ 
& \addpicv{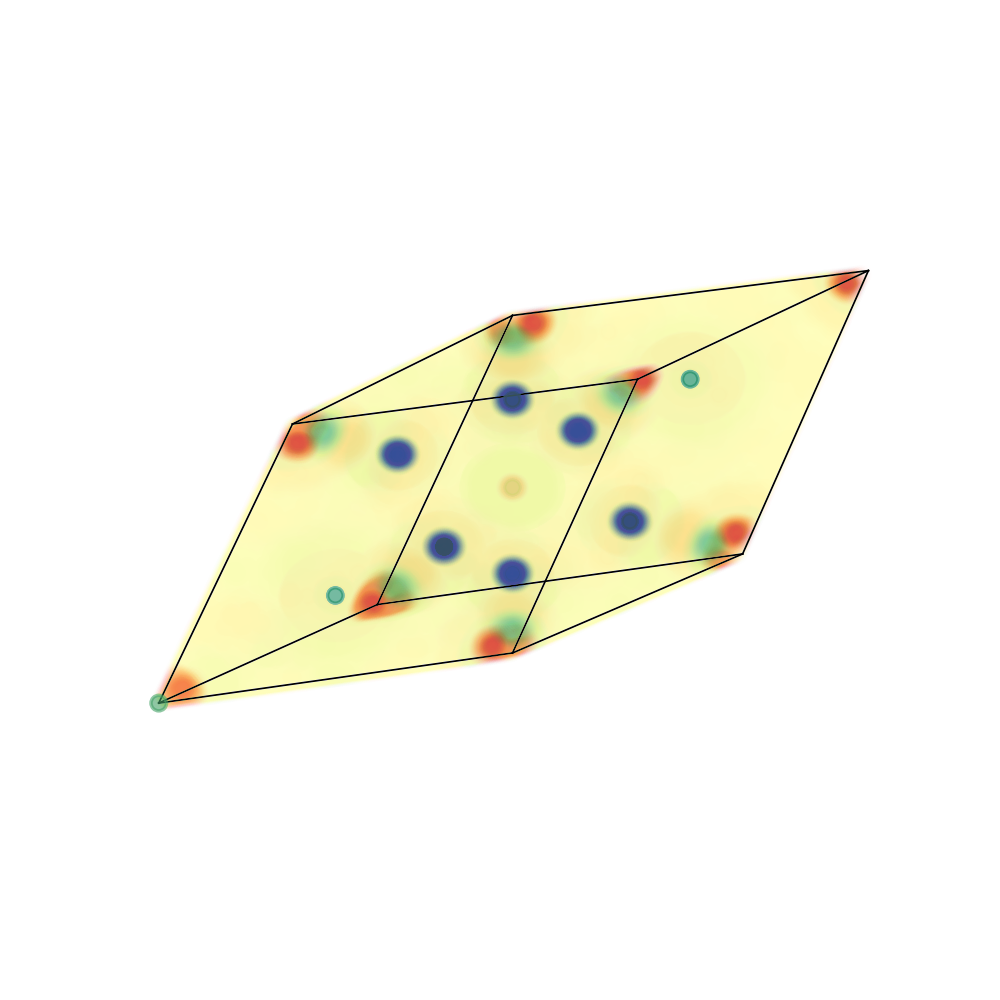}
& \addpicv{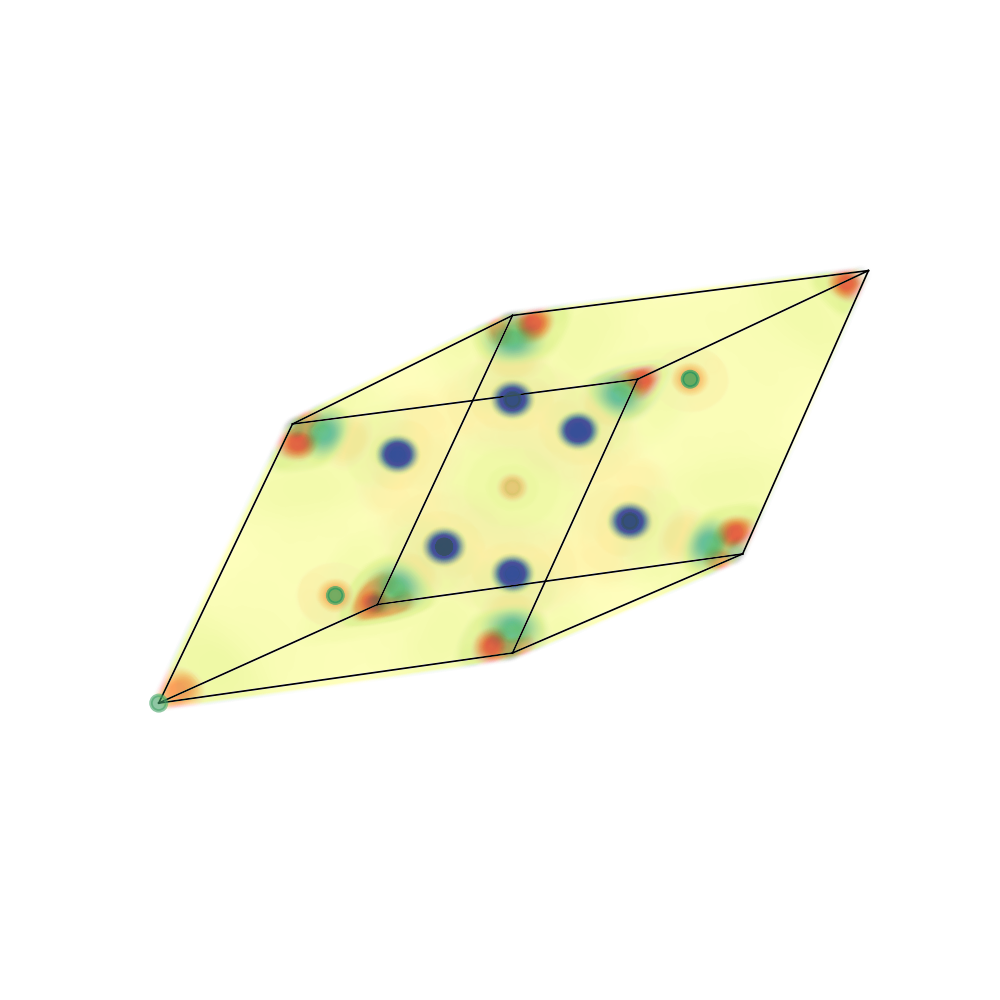}
& ~ 
& \addpich{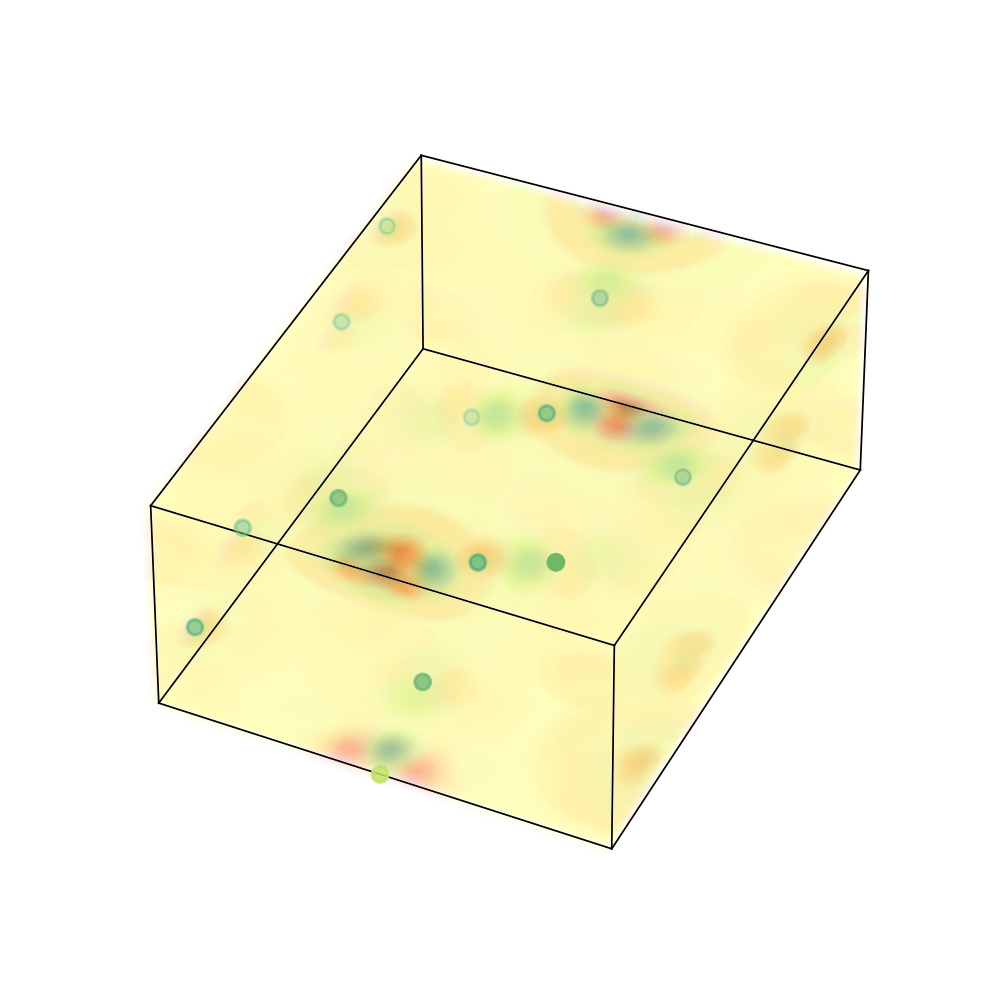}
& \addpich{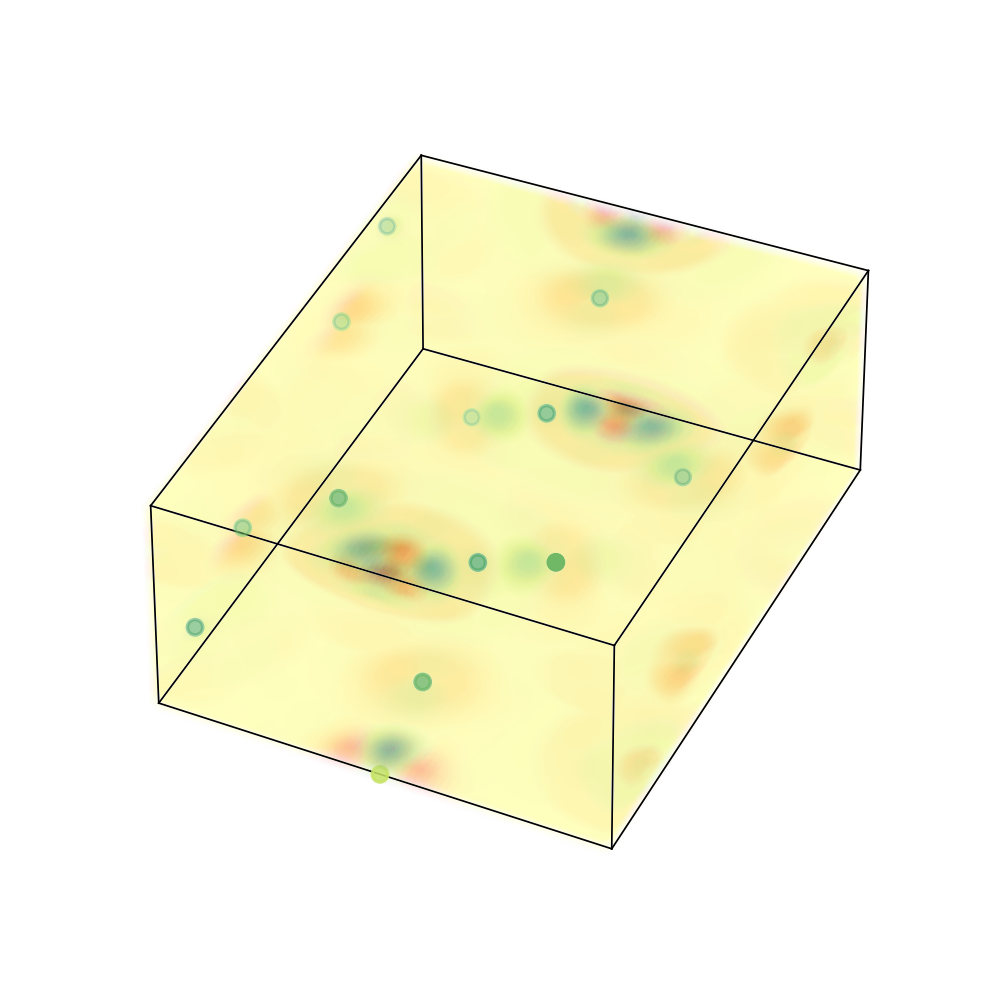}\\
\midrule
NMAE &~& \textbf{0.56} & 0.60 &~&\textbf{4.00}&5.89 &~&\textbf{4.59} & 5.15\\
\bottomrule 
\end{tabular}
}
\caption{\textbf{Visualization of ground truth (GT), density prediction, and error in NMAE (\%).} For electron density prediction (Pred.) on QM9 and MP datasets, red and blue colors indicate higher electron density, respectively. For the error (Error), lighter colors indicates lower error.}
\vspace{-1em}
\label{fig:vis}
\end{figure*} 



\textbf{High frequency masking.} To promote the synergy between the PW and GTO-based predictions, we employ high frequency mask (HFM), which constrains the PW prediction to low frequency region that is distant from atoms. To this end, we propose to mask the PW-based prediction near the atom as follows:
\begin{equation}
    a(\vx) = 
    \begin{cases} 1 \quad\text{if } \min_{u\in\mathcal{V}}d(\vx,\vx_{u}) > r_{\text{mask}}, \\
    0 \quad \text{else}.
    \end{cases}
\end{equation}
Note that our masking introduces discontinuities near the cutoff boundary, which we empirically observe to be ignorable.\footnote{One could replace the masking function with an envelope function \citep{gasteiger2020directional} for the continuous cutoff.} This results in the following prediction:
\begin{equation*}
    \rho_{\text{PW}}(\vx_{q}) = a(\vx)\mathbf{f}_{q}^{(H)},
\end{equation*} 
where $\vx_{q}$ is the input query point. We provide ablation studies on the effectiveness of HFM in \cref{sec:ablation}. 

\subsection{GTO-based Prediction} 
Finally, we describe our prediction using the GTO basis set. Our philosophy mostly follows the multicentric expansion scheme of InfGCN \citep{cheng2023equivariant}. Note that we can design the GTO-based prediction to satisfy the periodic boundary condition by introducing a periodical graph which is used in \citet{Xie18}.

The GTO-based prediction $\rho_{\text{GTO}}$ is constructed from the linear combination of GTO basis set as follows:
\begin{equation}
    \rho_{\text{GTO}}(\vx) = \sum_{u\in\mathcal{V}}\sum_{n=1}^{N}\sum_{l=1}^{L}\sum_{m=-l}^{l}\hat{f}_{u, nlm}\phi_{nlm}(\vx, \vx_{u}),
\end{equation}
where $\phi_{nlm}(\vx, \vx_{u})$ is the GTO as described in \cref{eq:gtobasis} centered at each atom position $\vx_{u}$ for $u\in\mathcal{V}$. Furthermore, $\hat{f}_{u, nlm} \in \mathbb{R}$ is the atom-wise coefficient term constructed from our neural operator given the molecule $\mathcal{M}$. Finally, $N$ and $L$ are the truncated maximum degrees for the radial basis and the spherical harmonics.

To be specific, the atom-wise coefficient terms are computed by TFN-based layer 
and its implementation \cite{thomas2018tensor,geiger2022e3nn} to achieve the equivariant message passing. The spherical tensor feature $\hat {\mathbf {f}}_{u,nl}^{(h)}=\{\hat {f}_{u,nlm}^{(h)}: m \in \mathbb Z, -l \le m \le l  \}$ is given by the following updates:
\begin{align}
    \hat {\mathbf{f}}^{(h+1)}_{u,nl} &= \sum_{v \in \mathcal{N}_{\gV}(u)} 
    \sum_{k\ge 0} W^{(h)}_{nlk}(\vx_{v}-\vx_{u})\hat{\mathbf{f}}^{(h)}_{v, nk}, \\
    W^{(h)}_{nlk}(\vr)  &= 
    \sum^{k+1}_{J=\|k-l\|}\varphi_{nlk}^J(r) \sum^J_{m=-J} Y_{Jm}\left(\hat{r}\right) Q^{lk}_{Jm},
\end{align}
where $r = \lVert \vr\rVert$, $\hat{r} = \frac{\vr}{r}$, $\mathcal{N}_{\gV}(u)$ is the set of neighbors of atom $u$,  $Q^{lk}_{Jm}\in \mathbb{R}^{(2l+2)\times(2k+1)}$ is the Clebsh-Gordan matrix, and $\varphi_{nlk}^J: \mathbb R^{+} \rightarrow \mathbb R$ is a learnable scalar function.

\section{Experiment}

\subsection{Experimental Setting}


\textbf{Baselines.}
The baseline methods of dataset include the GTO-based {InfGCN} \citep{cheng2023equivariant}, voxel based model CNN \citep{cciccek20163d}, interpolation-based models, i.e., DeepDFT \citep{jorgensen2020deepdft}, DeepDFT2  \citep{jorgensen2022equivariant}, EGNN \citep{satorras2021n}, {DimeNet} \citep{gasteiger2020directional}, {DimeNet++} \citep{gasteiger2020fast}, and neural operator based models, i.e., GNO \citep{li2020neural}, FNO \citep{li2020fourier}, and LNO \citep{lu2019deeponet}. 
We faithfully compare with the official numbers reported by the prior work \citep{cheng2023equivariant} when possible. We also outperform on another dataset, which we have not detailed here. We comprehensively compare in \cref{tab:full_table} of \cref{sec:full_table}. Our implementation of the GPWNO is published in the GitHub\footnote{\url{https://github.com/seongsukim-ml/GPWNO}}.

\textbf{Evaluation metric.} 
For evaluation, we follow prior works~\citep{jorgensen2020deepdft, cheng2023equivariant} in using the normalized mean absolute error~(NMAE):
\begin{equation}
    \operatorname{NMAE}(\rho, \rho_{\text{pred}}) = \frac{\int_{\R^3} \|\rho(\vx)-\rho_{\text{pred}}(\vx)\|\text{d}\vx}{\int_{\R^3} \|\rho(\vx)\|\text{d}\vx}.
\end{equation}
We provide more details regarding experiments setup, baselines, and evaluations in \cref{sec:train}.

\subsection{Aperiodic Materials}\label{sec:exp_aperiodic}
\textbf{Dataset.} For aperiodic molecules, following the prior work~\citep{cheng2023equivariant}, we consider electron density prediction for the QM9 dataset \citep{ruddigkeit2012enumeration, ramakrishnan2014quantum} and the molecular dynamics (MD) dataset \citep{bogojeski2018efficient}. Specifically, the electron density dataset of the QM9 was provided by \citet{Jørgensen_Bhowmik_2022}, and the MD was provided by \citet{brockherde2017bypassing,bogojeski2020quantum}. The MD dataset is further categorized into six smaller datasets corresponding to six types of molecules, i.e., ethanol, benzene, phenol, resorcinol, ethane, and malonaldehyde (MDA). Each smaller dataset contains snapshots of different geometries of a molecule. More details regarding the train and dataset are in the \cref{sec:train}.



\begin{figure}[t]
    \centering
    \includegraphics[width=\linewidth]{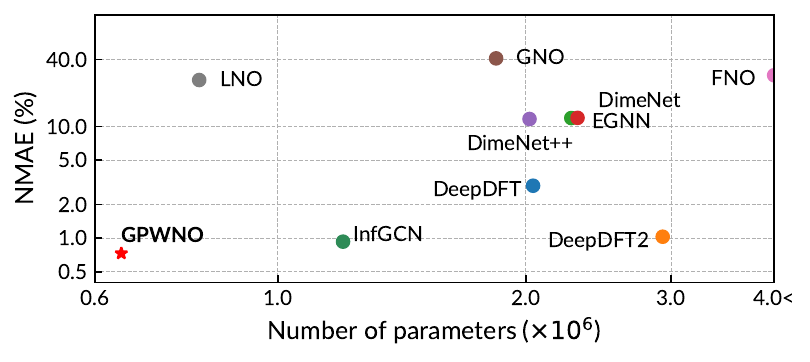}
    \caption{\textbf{Comparison between the baselines on QM9 (lower left is better).}}
    \label{fig:Params}
   \vspace{-.2in}
\end{figure}

\textbf{Quantitative results.} We present our experimental results in Table \ref{tab:aperiodic}. The results demonstrate that our algorithm significantly outperforms existing baselines. In particular, in the QM9 and MD datasets, our approach substantially improves over the second-best model, i.e., InfGCN, by approximately reducing the NMAE by 20\% in QM9 and 50\% in MD. One can also observe that FNO performs poorly despite its similarity to our GPWNO in using Fourier transformation as an intermediate layer. We hypothesize that the FNO cannot express the high-frequency details in the true electron density due to inefficient memory usage.


We also depict the tradeoff between the number of parameters and performance for the QM9 dataset in \cref{fig:Params}, where one can observe how GPWNO outperforms the considered baselines despite using the smallest number of parameters. The corresponding numbers are reported in \cref{sec:params}.


\textbf{Qualitative result.} We additionally visualize the predicted densities for the QM9 dataset in \cref{fig:vis}. Here, the performance of InfGCN is worse than GPWNO. 

\begin{table*}[t]
    \centering
    \caption{\textbf{Evaluation of GPWNO for periodic materials.} NMAE (\%) in the MP dataset, categorized by the seven crystal family and their combinations. The best number is highlighted in \textbf{bold}. Each number is averaged over three runs.}
    \begin{tabular}{lcccccccc}
    \toprule
     Model & Mixed & Triclinic & Monoclinic & Orthorhombic & Tetragonal & Trigonal & Hexagonal & Cubic \\
    \midrule
    DeepDFT & 11.50 & 32.33 & 50.54 & 22.30 & 37.68 & 23.89 & 18.61 & 27.30 \\
    DeepDFT2 & 15.11 & 30.20 & 49.74 & 20.50 & 41.21 & 22.13 & 12.87 & 27.63 \\
    InfGCN & \phantom{0}5.35 & \phantom{0}4.36 & \phantom{0}4.63 & \phantom{0}5.06 & \phantom{0}5.02 & \phantom{0}5.21 & \phantom{0}4.96 & \phantom{0}4.93 \\
    \midrule
    \textbf{GPWNO} & \phantom{0}\textbf{4.84} & \phantom{0}\textbf{3.92} & \phantom{0}\textbf{4.44} & \phantom{0}\textbf{4.73} & \phantom{0}\textbf{4.52} & \phantom{0}\textbf{4.92} & \phantom{0}\textbf{4.55} & \phantom{0}\textbf{4.32}  \\
    \bottomrule
    \end{tabular}
    \label{tab:periodic}
\end{table*}

\setlength{\belowcaptionskip}{0pt}
 \begin{table*}[t]
    \centering
    
    \caption{\textbf{Ablation study on MD and MP.} Performance in NMAE (\%) for the models using one or more of our algorithmic components, i.e., plane wave (PW), Gaussian-type orbital (GTO), and high-frequency masking (HFM), are reported. The best number is highlighted in \textbf{bold}. Each number is averaged over three runs.}
    \begin{tabular}{ccccccccc}
    \toprule
    \multicolumn{3}{c}{Model}& \multicolumn{3}{c}{MD} & \multicolumn{3}{c}{MP} \\
    \cmidrule(lr){1-3}\cmidrule(lr){4-6}\cmidrule(lr){7-9} PW & GTO & HFM & Benzene & Phenol & Resorcinol &  Mixed  & Triclinic & Cubic \\
    \midrule
    {\color{green}\cmark} & {\color{red}\xmark} & {\color{red}\xmark} & 6.92 & 8.27 & 8.12 & 48.69 & 49.20 & 48.12\\
    {\color{red}\xmark} & {\color{green}\cmark} & {\color{red}\xmark} & 2.90 & 3.24 & 2.96 & \phantom{0}5.35 & \phantom{0}4.36 & \phantom{0}4.93 \\
    {\color{green}\cmark} & {\color{green}\cmark} & {\color{red}\xmark} & {2.62} & {2.83} & {2.77} & \phantom{0}{5.00} & \phantom{0}3.96  & \phantom{0}4.62\\
     {\color{green}\cmark}& {\color{green}\cmark} &{\color{green}\cmark} & \textbf{2.45} & \textbf{2.68} &\textbf{2.73} & \phantom{0}\textbf{4.84} & \phantom{0}\textbf{3.92} & \phantom{0}\textbf{4.32}  \\
    \bottomrule
    \end{tabular}
    \label{tab:ablation}
    \vspace{-0.5em}
\end{table*}

\subsection{Periodic Systems}\label{sec:exp_periodic}
\textbf{Dataset.} To evaluate our model on periodic systems with diverse lattice structures, we newly conduct a benchmark from the materials project  \citep[MP]{jain2013commentary, shen2021representationindependent}. Precisely, we categorize the electron density for 117,535 molecules provided by \citet{shen2021representationindependent} into seven crystal families: triclinic, monoclinic, orthorhombic, tetragonal, trigonal, hexagonal, and cubic. Additionally, we conduct an extra dataset as a mixture of these crystal families. The size of the datasets ranges from 7,681 to 26,081. We use 1,000 or 500 samples for the test and validation, depending on the dataset size. We compare this dataset with the baseline results obtained from running the official repository codes on the newly curated datasets.


\textbf{Quantitative results.} 
We report the experimental results in \cref{tab:periodic}. Here, one can observe how our algorithm significantly outperforms the baselines, regardless of the crystal family. We also note that DeepDFT and DeepDFT2 perform poorly for the dataset, despite using the official repository.

\textbf{Qualitative results.}
We provide visualization of the predicted densities on the MP dataset in \cref{fig:vis}. Here, one can observe how the overall error and near atom region error both are reduced on GPWNO compared to InfGCN.




\subsection{Ablation Study} \label{sec:ablation}
\begin{figure}[t]
    \centering
    \includegraphics[width=\linewidth]{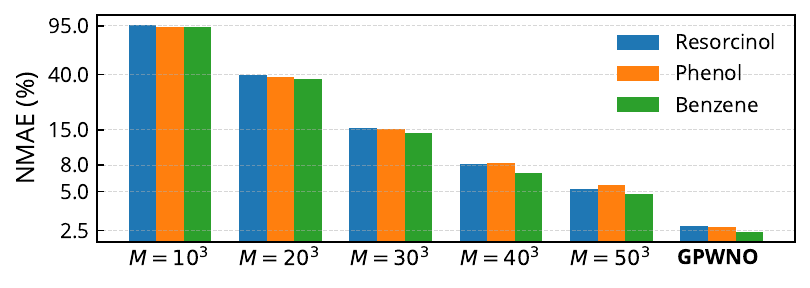}
    \caption{\textbf{Increasing number of probe nodes $M$ for PW-only prediction and comparing with GPWNO.}}
    \label{fig:Probes}
    \vspace{-1.7em}
\end{figure}

We conduct extensive ablation studies on the MD and MP to verify the effectiveness of our method. We provide the results in \cref{tab:ablation} and \cref{fig:Probes}. In \cref{tab:ablation}, we report the performance of our framework with and without each algorithmic component. In \cref{fig:Probes}, we check whether if introducing high-frequency components to the PW-only prediction eliminates the necessity of GTO basis. The detailed number of \cref{fig:Probes} is reported in \cref{tab:ablation2} of \cref{sec:full_table}.

\textbf{Independently evaluating PW and GTO basis.} In \cref{tab:ablation}, one can observe how PW- and GTO-based predictions perform without additional algorithmic components. Intriguingly, one can observe how the PW performs poorly without the GTO basis. We hypothesize this to be since the PW basis cannot capture high frequency components in the true signal. This is additionally supported by \cref{fig:Probes}, where PW demonstrates reasonable performance for the MD dataset when increasing the size of the PW basis. 

\setlength{\belowcaptionskip}{0pt}




\textbf{Synergy between PW and GTO basis.} One can verify the complementary behavior of PW and GTO basis from \cref{tab:ablation} since combining the two predictions consistently improve the performance. This verifies our idea that employing a dual representation is more effective than relying on a single basis alone, demonstrating the synergistic advantages of utilizing both representations.







\textbf{Effectiveness of HFM.} When comparing our algorithm with and without HFM in \cref{tab:ablation}, one can verify that HFM indeed enables the GTO-based prediction to concentrate on regions near atoms and PW-based prediction focus on regions far from atoms. Notably, HFM achieves this improvement in efficiency without introducing any additional model parameters.

\section{Conclusion}
In this work, we propose a new machine learning algorithm for electron density estimation. Through extensive experiments, we demonstrated the effectiveness of our model. While our experiments clearly demonstrate the superiority our GPWNO over the baselines, the evaluation metric is not fully aligned with our potential applications, e.g., errors in predicting the exchange-correlation energy. Since DFT computations required for this evaluation are expensive and rely on commercial licenses, we leave this for future work.

\newpage
\section*{Impact Statement}
Our framework can advance drug design and material discovery by accelerating molecular simulation. Such improvements often improve the quality of human life. However, they should also be approached with caution due to the potential misuse in the development of hazardous products.

\section*{Acknowledgement}
This work was partly supported by Institute of Information \& communications Technology Planning \& Evaluation (IITP) grant funded by the Korea government(MSIT) (No.RS-2019-II191906, Artificial Intelligence Graduate School Program(POSTECH)), the National Research Foundation of Korea(NRF) grant funded by the Korea government(MSIT) (No. 2022R1C1C1013366), and Basic Science Research Program through the National Research Foundation of Korea(NRF) funded by the Ministry of Education(2022R1A6A1A0305295413).

We thank Seungjin Kang, Suhwan Song, Byunggook Na, Yongdeok Kim, Minjong Lee, Seonghyun Park, Hyosoon Jang, Juwon Hwang, Kiyoung Seong, and Dongyeop Woo for providing helpful feedbacks and suggestions in preparing the early version of the manuscript.



\bibliography{999_reference}

\begin{thebibliography}{55}
\providecommand{\natexlab}[1]{#1}
\providecommand{\url}[1]{\texttt{#1}}
\expandafter\ifx\csname urlstyle\endcsname\relax
  \providecommand{\doi}[1]{doi: #1}\else
  \providecommand{\doi}{doi: \begingroup \urlstyle{rm}\Url}\fi

\bibitem[Adekoya et~al.(2022)Adekoya, Adekoya, Sadiku, Hamam, and Ray]{adekoya2022application}
Adekoya, O.~C., Adekoya, G.~J., Sadiku, E.~R., Hamam, Y., and Ray, S.~S.
\newblock Application of dft calculations in designing polymer-based drug delivery systems: An overview.
\newblock \emph{Pharmaceutics}, 14\penalty0 (9):\penalty0 1972, 2022.

\bibitem[Bachelet et~al.(1982)Bachelet, Hamann, and Schl{\"u}ter]{bachelet1982pseudopotentials}
Bachelet, G.~B., Hamann, D.~R., and Schl{\"u}ter, M.
\newblock Pseudopotentials that work: From h to pu.
\newblock \emph{Physical Review B}, 26\penalty0 (8):\penalty0 4199, 1982.

\bibitem[Bartlett(1981)]{bartlett1981many}
Bartlett, R.~J.
\newblock Many-body perturbation theory and coupled cluster theory for electron correlation in molecules.
\newblock \emph{Annual review of physical chemistry}, 32\penalty0 (1):\penalty0 359--401, 1981.

\bibitem[Bl{\"o}chl(1995)]{blochl1995electrostatic}
Bl{\"o}chl, P.
\newblock Electrostatic decoupling of periodic images of plane-wave-expanded densities and derived atomic point charges.
\newblock \emph{The Journal of chemical physics}, 103\penalty0 (17):\penalty0 7422--7428, 1995.

\bibitem[Bl{\"o}chl(1994)]{blochl1994projector}
Bl{\"o}chl, P.~E.
\newblock Projector augmented-wave method.
\newblock \emph{Physical review B}, 50\penalty0 (24):\penalty0 17953, 1994.

\bibitem[Bogojeski et~al.(2018)Bogojeski, Brockherde, Vogt-Maranto, Li, Tuckerman, Burke, and M{\"u}ller]{bogojeski2018efficient}
Bogojeski, M., Brockherde, F., Vogt-Maranto, L., Li, L., Tuckerman, M.~E., Burke, K., and M{\"u}ller, K.-R.
\newblock Efficient prediction of 3d electron densities using machine learning.
\newblock \emph{arXiv preprint arXiv:1811.06255}, 2018.

\bibitem[Bogojeski et~al.(2020)Bogojeski, Vogt-Maranto, Tuckerman, M{\"u}ller, and Burke]{bogojeski2020quantum}
Bogojeski, M., Vogt-Maranto, L., Tuckerman, M.~E., M{\"u}ller, K.-R., and Burke, K.
\newblock Quantum chemical accuracy from density functional approximations via machine learning.
\newblock \emph{Nature communications}, 11\penalty0 (1):\penalty0 5223, 2020.

\bibitem[Brockherde et~al.(2017)Brockherde, Vogt, Li, Tuckerman, Burke, and M{\"u}ller]{brockherde2017bypassing}
Brockherde, F., Vogt, L., Li, L., Tuckerman, M.~E., Burke, K., and M{\"u}ller, K.-R.
\newblock Bypassing the kohn-sham equations with machine learning.
\newblock \emph{Nature communications}, 8\penalty0 (1):\penalty0 872, 2017.

\bibitem[Chang et~al.(2021)Chang, Jiang, Wang, Zhao, Huang, and Chen]{chang2021lead}
Chang, J., Jiang, L., Wang, G., Zhao, W., Huang, Y., and Chen, H.
\newblock Lead-free perovskite compounds cssn 1- x ge x i 3- y br y explored for superior visible-light absorption.
\newblock \emph{Physical Chemistry Chemical Physics}, 23\penalty0 (26):\penalty0 14449--14456, 2021.

\bibitem[Cheng \& Peng(2023)Cheng and Peng]{cheng2023equivariant}
Cheng, C. and Peng, J.
\newblock Equivariant neural operator learning with graphon convolution.
\newblock In \emph{Thirty-seventh Conference on Neural Information Processing Systems}, 2023.
\newblock URL \url{https://openreview.net/forum?id=EjiA3uWpnc}.

\bibitem[{\c{C}}i{\c{c}}ek et~al.(2016){\c{C}}i{\c{c}}ek, Abdulkadir, Lienkamp, Brox, and Ronneberger]{cciccek20163d}
{\c{C}}i{\c{c}}ek, {\"O}., Abdulkadir, A., Lienkamp, S.~S., Brox, T., and Ronneberger, O.
\newblock 3d u-net: learning dense volumetric segmentation from sparse annotation.
\newblock In \emph{Medical Image Computing and Computer-Assisted Intervention--MICCAI 2016: 19th International Conference, Athens, Greece, October 17-21, 2016, Proceedings, Part II 19}, pp.\  424--432. Springer, 2016.

\bibitem[Fabrizio et~al.(2019)Fabrizio, Grisafi, Meyer, Ceriotti, and Corminboeuf]{fabrizio2019electron}
Fabrizio, A., Grisafi, A., Meyer, B., Ceriotti, M., and Corminboeuf, C.
\newblock Electron density learning of non-covalent systems.
\newblock \emph{Chemical science}, 10\penalty0 (41):\penalty0 9424--9432, 2019.

\bibitem[Gasteiger et~al.(2020{\natexlab{a}})Gasteiger, Giri, Margraf, and G{\"u}nnemann]{gasteiger2020fast}
Gasteiger, J., Giri, S., Margraf, J.~T., and G{\"u}nnemann, S.
\newblock Fast and uncertainty-aware directional message passing for non-equilibrium molecules.
\newblock \emph{arXiv preprint arXiv:2011.14115}, 2020{\natexlab{a}}.

\bibitem[Gasteiger et~al.(2020{\natexlab{b}})Gasteiger, Gro{\ss}, and G{\"u}nnemann]{gasteiger2020directional}
Gasteiger, J., Gro{\ss}, J., and G{\"u}nnemann, S.
\newblock Directional message passing for molecular graphs.
\newblock 2020{\natexlab{b}}.
\newblock URL \url{https://openreview.net/forum?id=B1eWbxStPH}.

\bibitem[Geiger \& Smidt(2022)Geiger and Smidt]{geiger2022e3nn}
Geiger, M. and Smidt, T.
\newblock e3nn: Euclidean neural networks.
\newblock \emph{arXiv preprint arXiv:2207.09453}, 2022.
\newblock URL \url{https://arxiv.org/abs/2207.09453}.

\bibitem[Giannozzi et~al.(2017)Giannozzi, Andreussi, Brumme, Bunau, Nardelli, Calandra, Car, Cavazzoni, Ceresoli, Cococcioni, et~al.]{giannozzi2017advanced}
Giannozzi, P., Andreussi, O., Brumme, T., Bunau, O., Nardelli, M.~B., Calandra, M., Car, R., Cavazzoni, C., Ceresoli, D., Cococcioni, M., et~al.
\newblock Advanced capabilities for materials modelling with quantum espresso.
\newblock \emph{Journal of physics: Condensed matter}, 29\penalty0 (46):\penalty0 465901, 2017.

\bibitem[Gill(1994)]{gill1994molecular}
Gill, P.~M.
\newblock Molecular integrals over gaussian basis functions.
\newblock In \emph{Advances in quantum chemistry}, volume~25, pp.\  141--205. Elsevier, 1994.

\bibitem[Gonze et~al.(2009)Gonze, Amadon, Anglade, Beuken, Bottin, Boulanger, Bruneval, Caliste, Caracas, C{\^o}t{\'e}, et~al.]{gonze2009abinit}
Gonze, X., Amadon, B., Anglade, P.-M., Beuken, J.-M., Bottin, F., Boulanger, P., Bruneval, F., Caliste, D., Caracas, R., C{\^o}t{\'e}, M., et~al.
\newblock Abinit: First-principles approach to material and nanosystem properties.
\newblock \emph{Computer Physics Communications}, 180\penalty0 (12):\penalty0 2582--2615, 2009.

\bibitem[Grisafi et~al.(2018)Grisafi, Fabrizio, Meyer, Wilkins, Corminboeuf, and Ceriotti]{grisafi2018transferable}
Grisafi, A., Fabrizio, A., Meyer, B., Wilkins, D.~M., Corminboeuf, C., and Ceriotti, M.
\newblock Transferable machine-learning model of the electron density.
\newblock \emph{ACS central science}, 5\penalty0 (1):\penalty0 57--64, 2018.

\bibitem[Hafner(2008)]{hafner2008ab}
Hafner, J.
\newblock Ab-initio simulations of materials using vasp: Density-functional theory and beyond.
\newblock \emph{Journal of computational chemistry}, 29\penalty0 (13):\penalty0 2044--2078, 2008.

\bibitem[Harrison(1966)]{harrison1966pseudopotentials}
Harrison, W.
\newblock \emph{Pseudopotentials in the Theory of Metals}.
\newblock Frontiers in Physics : a lecture note and reprint series. W.A. Benjamin, 1966.
\newblock ISBN 9780805337310.
\newblock URL \url{https://books.google.co.kr/books?id=XV8sAAAAYAAJ}.

\bibitem[Hehre et~al.(1972)Hehre, Ditchfield, and Pople]{hehre1972self}
Hehre, W.~J., Ditchfield, R., and Pople, J.~A.
\newblock Self—consistent molecular orbital methods. xii. further extensions of gaussian—type basis sets for use in molecular orbital studies of organic molecules.
\newblock \emph{The Journal of Chemical Physics}, 56\penalty0 (5):\penalty0 2257--2261, 1972.

\bibitem[Hendrycks \& Gimpel(2016)Hendrycks and Gimpel]{hendrycks2016gaussian}
Hendrycks, D. and Gimpel, K.
\newblock Gaussian error linear units (gelus).
\newblock \emph{arXiv preprint arXiv:1606.08415}, 2016.

\bibitem[Jain et~al.(2013)Jain, Ong, Hautier, Chen, Richards, Dacek, Cholia, Gunter, Skinner, Ceder, et~al.]{jain2013commentary}
Jain, A., Ong, S.~P., Hautier, G., Chen, W., Richards, W.~D., Dacek, S., Cholia, S., Gunter, D., Skinner, D., Ceder, G., et~al.
\newblock Commentary: The materials project: A materials genome approach to accelerating materials innovation.
\newblock \emph{APL materials}, 1\penalty0 (1), 2013.

\bibitem[J{\o}rgensen \& Bhowmik(2020)J{\o}rgensen and Bhowmik]{jorgensen2020deepdft}
J{\o}rgensen, P.~B. and Bhowmik, A.
\newblock Deepdft: Neural message passing network for accurate charge density prediction.
\newblock \emph{arXiv preprint arXiv:2011.03346}, 2020.

\bibitem[J{\o}rgensen \& Bhowmik(2022)J{\o}rgensen and Bhowmik]{jorgensen2022equivariant}
J{\o}rgensen, P.~B. and Bhowmik, A.
\newblock Equivariant graph neural networks for fast electron density estimation of molecules, liquids, and solids.
\newblock \emph{npj Computational Materials}, 8\penalty0 (1):\penalty0 183, 2022.

\bibitem[Jørgensen \& Bhowmik(2022)Jørgensen and Bhowmik]{Jørgensen_Bhowmik_2022}
Jørgensen, P.~B. and Bhowmik, A.
\newblock Equivariant graph neural networks for fast electron density estimation of molecules, liquids, and solids.
\newblock \emph{npj Computational Materials}, 8\penalty0 (1), 2022.
\newblock \doi{10.1038/s41524-022-00863-y}.

\bibitem[Kittel(2005)]{kittel2005introduction}
Kittel, C.
\newblock \emph{Introduction to solid state physics}.
\newblock John Wiley \& sons, inc, 2005.

\bibitem[Koelling \& Arbman(1975)Koelling and Arbman]{koelling1975use}
Koelling, D. and Arbman, G.
\newblock Use of energy derivative of the radial solution in an augmented plane wave method: application to copper.
\newblock \emph{Journal of Physics F: Metal Physics}, 5\penalty0 (11):\penalty0 2041, 1975.

\bibitem[Kohn \& Sham(1965)Kohn and Sham]{kohn1965self}
Kohn, W. and Sham, L.~J.
\newblock Self-consistent equations including exchange and correlation effects.
\newblock \emph{Physical review}, 140\penalty0 (4A):\penalty0 A1133, 1965.

\bibitem[Koker et~al.(2023)Koker, Quigley, Taw, Tibbetts, and Li]{koker2023higher}
Koker, T., Quigley, K., Taw, E., Tibbetts, K., and Li, L.
\newblock Higher-order equivariant neural networks for charge density prediction in materials.
\newblock \emph{arXiv preprint arXiv:2312.05388}, 2023.

\bibitem[Kondrakov et~al.(2017)Kondrakov, Ge{\ss}wein, Galdina, De~Biasi, Meded, Filatova, Schumacher, Wenzel, Hartmann, Brezesinski, et~al.]{kondrakov2017charge}
Kondrakov, A.~O., Ge{\ss}wein, H., Galdina, K., De~Biasi, L., Meded, V., Filatova, E.~O., Schumacher, G., Wenzel, W., Hartmann, P., Brezesinski, T., et~al.
\newblock Charge-transfer-induced lattice collapse in ni-rich ncm cathode materials during delithiation.
\newblock \emph{The journal of physical chemistry C}, 121\penalty0 (44):\penalty0 24381--24388, 2017.

\bibitem[Kosmala et~al.(2023)Kosmala, Gasteiger, Gao, and G{\"u}nnemann]{kosmala2023ewaldbased}
Kosmala, A., Gasteiger, J., Gao, N., and G{\"u}nnemann, S.
\newblock Ewald-based long-range message passing for molecular graphs.
\newblock In \emph{Proceedings of the 40th International Conference on Machine Learning}, pp.\  17544--17563, 2023.

\bibitem[Kovachki et~al.(2023)Kovachki, Li, Liu, Azizzadenesheli, Bhattacharya, Stuart, and Anandkumar]{kovachki2023neural}
Kovachki, N., Li, Z., Liu, B., Azizzadenesheli, K., Bhattacharya, K., Stuart, A., and Anandkumar, A.
\newblock Neural operator: Learning maps between function spaces, 2023.

\bibitem[Landau \& Binder(2021)Landau and Binder]{landau2021guide}
Landau, D. and Binder, K.
\newblock \emph{A guide to Monte Carlo simulations in statistical physics}.
\newblock Cambridge university press, 2021.

\bibitem[Li et~al.(2020{\natexlab{a}})Li, Kovachki, Azizzadenesheli, Liu, Bhattacharya, Stuart, and Anandkumar]{li2020fourier}
Li, Z., Kovachki, N., Azizzadenesheli, K., Liu, B., Bhattacharya, K., Stuart, A., and Anandkumar, A.
\newblock Fourier neural operator for parametric partial differential equations.
\newblock \emph{arXiv preprint arXiv:2010.08895}, 2020{\natexlab{a}}.

\bibitem[Li et~al.(2020{\natexlab{b}})Li, Kovachki, Azizzadenesheli, Liu, Bhattacharya, Stuart, and Anandkumar]{li2020neural}
Li, Z., Kovachki, N., Azizzadenesheli, K., Liu, B., Bhattacharya, K., Stuart, A., and Anandkumar, A.
\newblock Neural operator: Graph kernel network for partial differential equations.
\newblock \emph{arXiv preprint arXiv:2003.03485}, 2020{\natexlab{b}}.

\bibitem[Li et~al.(2023)Li, Kovachki, Choy, Li, Kossaifi, Otta, Nabian, Stadler, Hundt, Azizzadenesheli, et~al.]{li2023geometry}
Li, Z., Kovachki, N.~B., Choy, C., Li, B., Kossaifi, J., Otta, S.~P., Nabian, M.~A., Stadler, M., Hundt, C., Azizzadenesheli, K., et~al.
\newblock Geometry-informed neural operator for large-scale 3d pdes.
\newblock In \emph{Thirty-seventh Conference on Neural Information Processing Systems}, 2023.

\bibitem[Lippert et~al.(1997)Lippert, PARRINELLO, and MICHELE]{lippert1997hybrid}
Lippert, B.~G., PARRINELLO, J.~H., and MICHELE.
\newblock A hybrid gaussian and plane wave density functional scheme.
\newblock \emph{Molecular Physics}, 92\penalty0 (3):\penalty0 477--488, 1997.

\bibitem[Lu et~al.(2019)Lu, Jin, and Karniadakis]{lu2019deeponet}
Lu, L., Jin, P., and Karniadakis, G.~E.
\newblock Deeponet: Learning nonlinear operators for identifying differential equations based on the universal approximation theorem of operators.
\newblock \emph{arXiv preprint arXiv:1910.03193}, 2019.

\bibitem[Mardirossian et~al.(2018)Mardirossian, McClain, and Chan]{mardirossian2018lowering}
Mardirossian, N., McClain, J.~D., and Chan, G.~K.
\newblock Lowering of the complexity of quantum chemistry methods by choice of representation.
\newblock \emph{The Journal of chemical physics}, 148\penalty0 (4), 2018.

\bibitem[Martin(2004)]{martin_2004}
Martin, R.~M.
\newblock \emph{Electronic Structure: Basic Theory and Practical Methods}.
\newblock Cambridge University Press, 2004.
\newblock \doi{10.1017/CBO9780511805769}.

\bibitem[Ramakrishnan et~al.(2014)Ramakrishnan, Dral, Rupp, and Von~Lilienfeld]{ramakrishnan2014quantum}
Ramakrishnan, R., Dral, P.~O., Rupp, M., and Von~Lilienfeld, O.~A.
\newblock Quantum chemistry structures and properties of 134 kilo molecules.
\newblock \emph{Scientific data}, 1\penalty0 (1):\penalty0 1--7, 2014.

\bibitem[Ruddigkeit et~al.(2012)Ruddigkeit, Van~Deursen, Blum, and Reymond]{ruddigkeit2012enumeration}
Ruddigkeit, L., Van~Deursen, R., Blum, L.~C., and Reymond, J.-L.
\newblock Enumeration of 166 billion organic small molecules in the chemical universe database gdb-17.
\newblock \emph{Journal of chemical information and modeling}, 52\penalty0 (11):\penalty0 2864--2875, 2012.

\bibitem[Satorras et~al.(2021)Satorras, Hoogeboom, and Welling]{satorras2021n}
Satorras, V.~G., Hoogeboom, E., and Welling, M.
\newblock E (n) equivariant graph neural networks.
\newblock In \emph{International conference on machine learning}, pp.\  9323--9332. PMLR, 2021.

\bibitem[Sch{\"u}tt et~al.(2021)Sch{\"u}tt, Unke, and Gastegger]{schutt2021equivariant}
Sch{\"u}tt, K., Unke, O., and Gastegger, M.
\newblock Equivariant message passing for the prediction of tensorial properties and molecular spectra.
\newblock In \emph{International Conference on Machine Learning}, pp.\  9377--9388. PMLR, 2021.

\bibitem[Shen et~al.(2021)Shen, Munro, Horton, Huck, Dwaraknath, and Persson]{shen2021representationindependent}
Shen, J.-X., Munro, J.~M., Horton, M.~K., Huck, P., Dwaraknath, S., and Persson, K.~A.
\newblock A representation-independent electronic charge density database for crystalline materials, 2021.

\bibitem[Sinitskiy \& Pande(2018)Sinitskiy and Pande]{sinitskiy2018deep}
Sinitskiy, A.~V. and Pande, V.~S.
\newblock Deep neural network computes electron densities and energies of a large set of organic molecules faster than density functional theory (dft).
\newblock \emph{arXiv preprint arXiv:1809.02723}, 2018.

\bibitem[Sun et~al.(2023)Sun, Ross, Zhu, and Azizzadenesheli]{sun2023phase}
Sun, H., Ross, Z.~E., Zhu, W., and Azizzadenesheli, K.
\newblock Phase neural operator for multi-station picking of seismic arrivals.
\newblock \emph{Geophysical Research Letters}, 50\penalty0 (24):\penalty0 e2023GL106434, 2023.

\bibitem[Sun et~al.(2017)Sun, Berkelbach, McClain, and Chan]{sun2017gaussian}
Sun, Q., Berkelbach, T.~C., McClain, J.~D., and Chan, G.~K.
\newblock Gaussian and plane-wave mixed density fitting for periodic systems.
\newblock \emph{The Journal of chemical physics}, 147\penalty0 (16), 2017.

\bibitem[Thomas et~al.(2018)Thomas, Smidt, Kearnes, Yang, Li, Kohlhoff, and Riley]{thomas2018tensor}
Thomas, N., Smidt, T., Kearnes, S., Yang, L., Li, L., Kohlhoff, K., and Riley, P.
\newblock Tensor field networks: Rotation-and translation-equivariant neural networks for 3d point clouds.
\newblock \emph{arXiv preprint arXiv:1802.08219}, 2018.

\bibitem[Tran et~al.(2023)Tran, Mathews, Xie, and Ong]{tran2023factorized}
Tran, A., Mathews, A., Xie, L., and Ong, C.~S.
\newblock Factorized fourier neural operators.
\newblock In \emph{The Eleventh International Conference on Learning Representations}, 2023.
\newblock URL \url{https://openreview.net/forum?id=tmIiMPl4IPa}.

\bibitem[Ulian et~al.(2013)Ulian, Tosoni, and Valdr{\`e}]{ulian2013comparison}
Ulian, G., Tosoni, S., and Valdr{\`e}, G.
\newblock Comparison between gaussian-type orbitals and plane wave ab initio density functional theory modeling of layer silicates: Talc [mg3si4o10 (oh) 2] as model system.
\newblock \emph{The Journal of chemical physics}, 139\penalty0 (20), 2013.

\bibitem[Wang et~al.(2022)Wang, Choudhary, Liu, Hu, and Hu]{wang2022large}
Wang, F.~Q., Choudhary, K., Liu, Y., Hu, J., and Hu, M.
\newblock Large scale dataset of real space electronic charge density of cubic inorganic materials from density functional theory (dft) calculations.
\newblock \emph{Scientific Data}, 9\penalty0 (1):\penalty0 59, 2022.

\bibitem[Xie \& Grossman(2018)Xie and Grossman]{Xie18}
Xie, T. and Grossman, J.~C.
\newblock Crystal graph convolutional neural networks for an accurate and interpretable prediction of material properties.
\newblock \emph{Phys. Rev. Lett.}, 120:\penalty0 145301, Apr 2018.
\newblock \doi{10.1103/PhysRevLett.120.145301}.
\newblock URL \url{https://link.aps.org/doi/10.1103/PhysRevLett.120.145301}.

\end{thebibliography}
\bibliographystyle{icml2024}

\newpage
\appendix
\onecolumn


\section{Overall Architecture}\label{sec:equivariance}\label{sec:arch}
\begin{figure*}[h]
    \centering
    \includegraphics[width=0.9\linewidth]{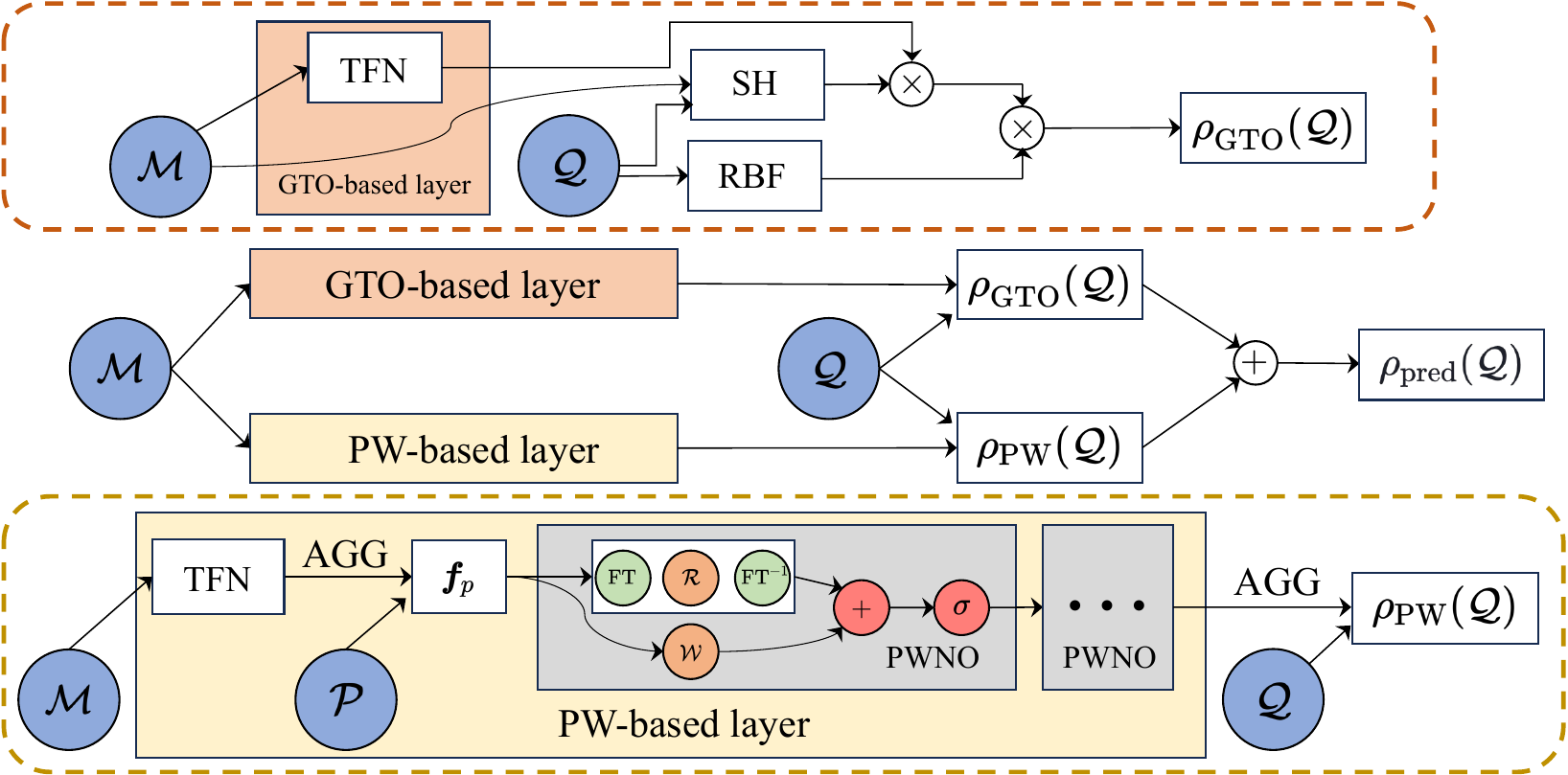}
    \caption{\textbf{Architecture of Gaussian plane-wave neural operator (GPWNO).} Starting from molecule $\gM=(\gX,\gA,\mL)$ with the atomic positions $\gX$, atomic features $\gA$ and the lattice vectors $\mL$, GPWNO aims to predict a scalar field $\rho_{\text{pred}}$ for given query points $\gQ \subset \R^3$. Notably, we decompose the target $\rho_{\text{pred}}$ into two types of the bases: Gaussian-type orbital (GTO) basis and  Plane wave (PW) basis. \textbf{(top)} GTO basis layer predict the $\rho_{\text{GTO}}$ by the direct coefficients learning of spherical harmonics. \textbf{(bottom)} PW basis layer predict the signals of reciprocal space by using discretized voxel points $\gP$ and the FNO layers, and predict the $\rho_{\text{PW}}$ the message aggregation from voxel points $\gP$.}
    \label{fig:method_app}
\end{figure*}

In \cref{fig:method_app}, we provide a detailed description of our model. 

\newpage 

\section{Signal Initialization by Tensor Field Network}\label{sec:siginit}

Here, we describe our TFN-based architecture for initializing a signal as an input to our PW neural operator. For a given molecule $\gM=(\gX,\gA,\mL)$, we first initialize the atomic feature $\vf_{u}^{(0)}$ of atom $u \in \gV$ by the linear-embedding layer that takes input as the one-hot encoding of the atomic number of each atom. Then, we update the atomic feature with the TFN layer by treating $\vf_{u}^{(0)} \in \mathbb R^C$ as a multi-channel rank-0 tensor feature of dimension $C$. i.e., $\vf_{u}^{(0)}=\{f^{1}_{u,00},\ldots,f^{C}_{u,00}\}$. Specifically, each embedding layer updates the spherical node feature ${\vf}_{u,l}^{(h)}=\{\hat {f}_{u,lm}^{(h)}: m \in \mathbb Z, -l \le m \le l  \}$ from the near atoms spherical node feature ${\vf}_{v,l}^{(h)}$ by tensor product:
\begin{align}
    \label{eq:node_embedding}
    \hat {\mathbf{f}}^{(h+1)}_{u,l} &= \sum_{v \in \mathcal{N}(u)} 
    \sum_{k\ge 0} W^{(h)}_{lk}(\vx_{v}-\vx_{u})\hat{\mathbf{f}}^{(h)}_{v, k}, \\
    W^{(h)}_{lk}(\vr)  &= 
    \sum^{k+1}_{J=\|k-l\|}\varphi_{lk}^J(\lVert \vr\rVert) \sum^J_{m=-J} Y_{Jm}\left(\frac{\vr}{\lVert \vr \rVert}\right) Q^{lk}_{Jm}.
\end{align}
Here, $\mathcal{N}(u)$ is the set of neighbors of atom $u$,  $Q^{lk}_{Jm}\in \mathbb{R}^{(2l+2)\times(2k+1)}$ is the Clebsh-Gordan matrix, and $\varphi_{nlk}^J: \mathbb R^+ \rightarrow \mathbb R$ is a learnable scalar function. At the final embedding layer $H$, we set the output feature of the TFN as the multi-channel rank-0 tensor with dimension $K$. i.e., $\vf_{u}^{(H)} \in \mathbb R^K$. Finally, we use this $\vf_{u}^{(H)}$ as the input of the \cref{eq:probe1}.

\newpage
\section{Proof of Equivariance for the Lattice Construction on Aperiodic Molecule}\label{sec:eqproof}

We define the lattice $\boldsymbol L$ of the aperiodic molecule based on the principal component analysis (PCA). For a given molecule $\gM$, let $n$ is the number of the atoms and $\rmX$ is the atom positon matrix. i.e., $\rmX=[\vx_1,\ldots,\vx_n]^\top \in \mathbb R^{n\times 3}$. Then, let $\vt = \frac{1}{n}\sum_{u \in \gV} \vx_u \in \mathbb R^3$ be the centroid of the molecule $\rmX$ with $n$ atoms and $\rmC=(\rmX-\bold{1}\vt^\top)^\top(\rmX-\bold{1}^\top) \in \mathbb R^{3\times 3}$ be the covariance matrix computed after removing the centroid from $\rmX$. In the generic case, the eigenvalues of $\rmC$ satisfy $\lambda_1 < \lambda_2 < \lambda_3$. Let $\vv_1,\vv_2,\vv_3$ be the unit length corresponding eigenvectors. Then we define $\boldsymbol L=[\vv_1,\vv_2,\vv_3]$ Note that eigenvectors are orthogonal to each other, so $\boldsymbol L$ can be used to the cubical lattice vector.

Next, we will show this $\boldsymbol L$ is SE(3) equivariant, meaning that the rotation of the data naturally transforms the lattice vector, i.e, $\boldsymbol L' =[\vv_1\rmR^\top,\vv_2\rmR^\top,\vv_3\rmR^\top]$. Note that when we think about the direction of PCA eigenvector points out the maximal variance axis, it is natural that eigenvectors rotatated equivalently to the frame.

Let $(\rmR,\vk)\in$ SO(3); then the position of molecule are transformed to $\rmX'=\rmX\rmR+\bold{1}\vk^\top$. The centroid of the molecule is $\vt'=\rmR^\top\vt+\vk$, and the covariance matrix becomes
\begin{align*}
    \rmC'&=(\rmX'-\bold{1}\vt'^\top)^\top(\rmX'-\bold{1}\vt'^\top)\\
      &=(\rmX\rmR+\bold{1}\vk^\top-\bold{1}\vt^\top\rmR-\bold{1}\vk^\top)^\top(\rmX\rmR+\bold{1}\vk^\top-\bold{1}\vt^\top\rmR + \bold{1}\vk^\top)\\
      &=\rmR^\top(\rmX-\bold{1}\vt^\top)^\top(\rmX-\bold{1}\vt^\top)\rmR\\
      &=\rmR^\top \rmC\rmR
\end{align*}

By the definition of the eigen decomposition, the eigenvector $\vv$ of covaraince matrix $\rmC$ with eigenvalue $\lambda$ satisfies $\rmC\vv=\lambda\vv$. Similary, the transformed covariance matrix $\rmC'$ has the new eigenvector $\vv'=\rmR^\top \vv$. We show that $\vv'$ is the eigenvector of $\rmC'$ and its eigenvalue is $\lambda$:
\begin{align*}
\rmC'\vv'&=(\rmR^\top \rmC \rmR)(\rmR^\top \vv)\\
&=\rmR^\top \rmC \vv=\rmR^\top \lambda \vv\\
&= \lambda \vv'
\end{align*}
To this end, the lattice system of the rotated system is naturally rotated, that is $\boldsymbol L'=[\vv'_1,\vv'_2,\vv'_3]=[\vv_1\rmR^\top,\vv_2\rmR^\top,\vv_3\rmR^\top]$ as mentioned above.

\newpage
\section{Dataset and Hyperparameters}\label{sec:train}
In this section, we provide more details about our dataset and train configuration. All the dataset contains the molecule graph of each species and its voxelized 3D electron charge density data on its pre-defined grid. 

\textbf{QM9.} The dataset is provided by \citet{jorgensen2022equivariant,jorgensen2020deepdft} The electron density of the dataset are calculated by VASP. To be specific,  VASP with Perdew-Burke-Ernzerhof exchange correlation (PBE XC) functional and projector-augemented wave (PAW) is used.

\textbf{MD.} The dataset is provided by \citet{brockherde2017bypassing,bogojeski2018efficient,bogojeski2020quantum} and it is calculated by PBE XC functional with Quantum ESPRESSO using the PBE functional.

\textbf{Cubic2.} Cubic2 is not directly reported, since it shares the same materials structure with the cubic of the Materials Project. We report this to point out our model outperformed than previous works on this dataset, too. The data is provided by \citep{wang2022large} and calculated by the VASP with the PAW. The performance on Cubic2 is in \cref{tab:full_table}.

\textbf{Materials project.} This data is set of the periodic materials including cubic, hexagonal and so on. The detailed categorized condition is in \cref{tab:crystal_family}.

Common statistics of the QM9, Cubic2 and MD are summarized in \cref{tab:dataset_app}. MD dataset does not have a validation split, and the number of grids and grid lengths is for a single dimension so the number of voxels cubically with respect to it.

\begin{table}[h!]
    \centering
    \caption{\textbf{Statistics of the dataset.} We round up the number of grids and nodes to the nearest integer for readability.}
    \begin{tabular}{lccc}
    \toprule
         Dataset & {QM9}  & {Cubic2}  & {MD}\\
    \midrule
         train/val/test split & 123835/50/1600  & 14421/1000/1000 & 1000(2000)/500(400)\\
         min/mean/max \#grid & 40/88/160  & 32/94/448   & 20/20/20 \\
         min/mean/max \#node & 3/18/29 & 1/10/64  & 8/11/14 \\
         min/mean/max length (Bohr) & 4.00/8.65/15.83 & 1.78/5.82/26.20  & 20/20/20 \\
         \# node type & 5 & 84 & 3\\
    \bottomrule
    \end{tabular}
    \label{tab:dataset_app}
\end{table}

Additional statistics of the MP dataset are detailed in \cref{tab:mp_dataset_app}. The mixed dataset is composed of samples from each crystal family, with a distribution of 2000 for training, 150 for validation, and 150 for testing, resulting in a total of 14,000 training, 1,050 validation, and 1,050 testing samples. Note that the MP dataset may include non-cubical voxels, as its lattice structure is not confined to cubic forms. Differing from cubic structures, the generalized lattice types vary in length and grid dimensions along the $\va_1, \va_2,$ and $\va_3$ directions of the lattice vector. Each number of the grid binning and lattice length are denoted by "\# grid $\va_i$" and "length $\va_i$", respectivly. Accordingly, we provide statistics for each side, as well as the average binning of grid and length, which are represented as "\# grid avg", and "length avg". These averages are calculated using the cube root of the product of the values along $\va_1$, $\va_2$, and $\va_3$.

\begin{table}[h!]
\centering
\caption{\textbf{Dataset details of materials project.} We round up the number of grids to the nearest integer for readability. Each splitting of the numbers implies min, mean and max from the left to right. }
\resizebox{\textwidth}{!}{%
\begin{tabular}{lcccccccccc}
\toprule
& \textbf{Triclinic} & \textbf{Monoclinic} & \textbf{Orthorhombic} & \textbf{Tetragonal} & \textbf{Trigonal} & \textbf{Hexagonal} & \textbf{Cubic}  \\
\midrule
train/val/test split & 12843/1000/1000 & 26081/1000/1000 & 22115/1000/1000 & 12031/1000/1000 & 9651/500/500 & 7681/500/500 & 15133/1000/1000 \\
\# atom type & 83 & 86 & 84 & 85 & 84 & 87 & 88 \\
\# grid $\va_1$ & 36/103/1120 & 36/\phantom{0}99/972 & 28/\phantom{0}99/\phantom{0}420 & 24/101/\phantom{0}500 & 24/117/\phantom{0}576 & 28/\phantom{0}93/\phantom{0}280 & 28/90/180\\
\# grid $\va_2$ & 36/104/\phantom{0}540 & 36/118/756 & 28/100/\phantom{0}960 & 24/120/1120 & 20/112/\phantom{0}540 & 28/\phantom{0}98/\phantom{0}280 & 28/90/192 \\
\# grid $\va_3$ & 36/182/1512 & 48/162/896 & 16/126/1120 & 36/155/1120 & 36/156/1500 & 16/125/1440 & 28/90/240 \\
\# grid avg. & 47/115/\phantom{0}540 & 57/121/387 & 30/105/\phantom{0}420 & 38/119/\phantom{0}384 & 40/123/\phantom{0}384 & 33/\phantom{0}99/\phantom{0}178 & 28/90/184  \\
Length $\va_1$ (Bohr) & 2.36/\phantom{0}6.67/\phantom{0}74.23 & 2.30/\phantom{0}6.44/65.40 & 1.78/6.46/28.11 & 1.61/\phantom{0}6.60/32.62 & 1.52/7.64/38.79 & 1.71/6.01/18.30 & 1.78/5.85/12.10 \\
Length $\va_2$ (Bohr) & 2.36/\phantom{0}6.76/\phantom{0}34.99 & 2.36/\phantom{0}7.66/50.18 & 1.78/6.57/61.13 & 1.61/\phantom{0}7.85/74.88 & 1.28/7.34/36.35 & 1.71/6.33/18.30 & 1.78/5.85/12.80 \\
Length $\va_3$ (Bohr) & 2.49/11.95/101.39 & 2.94/10.58/59.56 & 0.98/8.20/75.07 & 2.34/10.13/74.43 & 2.46/10.17/100.46 & 0.98/8.16/95.06 & 1.78/5.86/15.92 \\
Length avg. (Bohr) & 3.10/\phantom{0}8.46/\phantom{0}35.94 & 3.78/\phantom{0}8.23/29.96 & 2.07/7.08/41.50 & 2.60/8.19/28.01 & 2.68/8.38/35.60 & 2.22/6.84/33.73 & 1.78/5.85/12.23 \\
\# node & 1/18/118 & 3/36/144 & 1/17/120 & 1/25/154 & 1/33/144 & 1/15/132 & 1/11/136 \\
\bottomrule
\end{tabular}
}%
\label{tab:mp_dataset_app}
\end{table}



\begin{table*}[!ht]
    \centering
    \caption{\textbf{The condition of crystal familiy.} Here, each $a, b, c$ represents the length of the corresponding lattice vector, i.e., $a = |\va_1|$, $b = |\va_2|$, and $c = |\va_3|$. Additionally, $\alpha$, $\beta$, and $\gamma$ denote the angles between $\va_2$ and $\va_3$, $\va_1$ and $\va_3$, and $\va_2$ and $\va_1$, respectively.  }
    \begin{tabular}{lcc}
    \toprule
    \multicolumn{1}{c}{\multirow{1}*{Crystal Famlily}} & Edge & Angle \\
    \midrule
     Triclinic      & $a\ne b \ne c$           & $\alpha \ne \beta \ne \gamma \ne 90^\circ$   \\
     Monoclinic     & $a\ne b \ne c$    & $\alpha = \gamma=90^\circ, \beta \ne 120^\circ$  \\
     Orthorhombic   & $a\ne b \ne c$    & $\alpha = \beta = \gamma = 90^\circ$   \\
     Tetragonal     & $a= b \ne c$      & $\alpha = \beta = \gamma = 90^\circ$  \\
     Trigonal       & $a= b = c$        & $\alpha = \beta = \gamma \ne 90^\circ$  \\
     Hexagonal      & $a= b \ne c$      & $\alpha = \beta = 90^\circ ,\gamma = 120^\circ$  \\
     Cubic          & $a = b = c$     & $\alpha = \beta = \gamma = 90^\circ$   \\
    \bottomrule
    \end{tabular}
    \label{tab:crystal_family}
\end{table*}
\newpage
~
\newpage
\section{Model and Training Details} 
\subsection{Model Details}

We provide more details about the proposed GWPNO model and the baseline.

\textbf{GPWNO.}\footnote{\url{https://github.com/seongsukim-ml/GPWNO}} Our implementation of GTO layer is inspired by the InfGCN. We used the spherical degree $l=3$ or $l=4$. The number of radio bases is settled $n=16$ with the Gaussian parameters starting at 0.5 (Bohr) and ending at 5.0. We used 3 layers of the GTO layers and the number of the trainable parameter of GTO layer is 223K when $l=3$ and 353K when $l=4$.

In the implementation of the PW layer, we set the spherical degree $l=3$ for embedding purposes. The number of probe nodes $M$ was configured to either 20 or 40. To enhance the scalability of the PW layer, we incorporated the approach outlined by \citet{tran2023factorized}. However, our empirical observations indicate that the performance of the model remains nearly unchanged with or without the use of \citet{tran2023factorized}, provided that the number of probe nodes remains consistent. Regarding the trainable parameters in the PW neural operator layer, the count stands at 19.3K for $M=20$ and increases to 47.0K when $M=40$.


\textbf{InfGCN.}\footnote{\url{https://github.com/ccr-cheng/infgcn-pytorch}} InfGCN is a GTO-based model designed to learn the direct coefficients of the GTO basis. We used the official code implementation of InfGCN by \citet{cheng2023equivariant}. 

\textbf{DeepDFT} and \textbf{DeepDFT2.}\footnote{\url{https://github.com/peterbjorgensen/DeepDFT}} DeepDFT is a GNN based model that uses message passing to the query node where the charge density is predicted. DeepDFT2 uses PaiNN \citep{schutt2021equivariant} as the GNN architecture. We used the official code implementation of DeepDFT and DeepDFT2 by \citet{jorgensen2020deepdft,jorgensen2022equivariant}.

\subsection{Training Details}
In this section, we outline our model specifications and those of the baseline models. Additionally, we detail the hyperparameters related to training and testing utilized in our experiments. As previously mentioned, we make direct comparisons with the official results reported in prior work by \citet{cheng2023equivariant} whenever possible.

\begin{table}[h!]
    \caption{\textbf{Hyperaparameter details}}
    \centering
    \resizebox{1.0\textwidth}{!}{%
    \begin{tabular}{llccccccccccc}
        \toprule
        Model & Dataset & \texttt{sp} & \texttt{cutoff} & \texttt{q.cutoff} & \texttt{p.cutoff} & $M$ & \texttt{num.grad}  & learning rate    & batch size & learning rate decay  &  query points \\
        \midrule
        \multirow{4}*{GPWNO}
        & QM9 & 4 & 3.0 & 1.5 & 0.75  & 40 &  80k  & 1e-3   & \multirow{4}*{32} & \multirow{4}*{0.5}  &  \multirow{4}*{1024} \\
        & MD & 3 & 3.0 & 1.5 & 0.75  & 40 & 4k  & 5e-3    &  &  &   \\
        & Cubic & 4 & 5.0 &2.5 & 1.25 & 20  & 10k  & 5e-3    &  &   &   \\
        & MP & 3 & 3.0 & 2.5 & 1.25  & 20 & 10k  & 5e-3  & &   &   \\
        \midrule
        \multirow{1}*{InfGCN}
        & MP & 4 &  3.0  & - & - & - & 10k  & 5e-3   & 32 &  0.5 & 1024  \\
        \midrule
        DeepDFT
        & \multirow{2}*{MP} & \multirow{2}*{4} &  \multirow{2}*{4.0}  & \multirow{2}*{-} & \multirow{2}*{-} & \multirow{2}*{-} & \multirow{2}*{10k}  & \multirow{2}*{3e-4}   & \multirow{2}*{32} & \multirow{2}*{0.5} & \multirow{2}*{1024}   \\
        DeepDFT2 & \\
        \bottomrule

    \end{tabular}
    \label{tab:hyper}
    }%
\end{table}

We list the training specifications and major hyperparameters of our experiments in \cref{tab:hyper}. Following the approach of \citet{cheng2023equivariant} and \citet{jorgensen2020deepdft}, we employ a sampling training scheme. In each training iteration, we sample multiple query points from the given data batch. For instance, we sample 1024 query points from a batch of 32 data points, resulting in a target size of $(32\times 1024)$ query points during training. For testing, we evaluate all samples in the test dataset to avoid randomness. Minibatch inference does not impact the results, so we segment the full grid into small sections, evaluate the points sequentially as mini-batch, and compute the normalized mean absolute error (NMAE) across the entire inference dataset.

We utilized the ADAM optimizer, and the learning rate and its decay also reported in \cref{tab:hyper}. In addition the number of the gradient step is also reported as \texttt{num.grad}.


Key hyperparameters in our model include the spherical harmonics (\texttt{sp}) order in the GTO layer and the radius cutoff distance for each message-passing operation. The \texttt{cutoff} parameter defines the neighborhood radius for the GTO layer. Additionally, the probe cutoff (\texttt{p.cutoff}) and query cutoff (\texttt{q.cutoff}) parameters specify the radius distances used in \cref{eq:probe1} and \cref{eq:query1}, respectively.

\newpage
\section{Additional Results}\label{sec:full_table}\label{sec:params}
We provide the additional results that is not included in the main context for the space.

\subsection{Additional Comparison}
We provide the performance on another data set that is not reported in the main context. For QM9, we also evaluate the performance on the randomly rotated test set. We observe that our GPWNO also outperform than previous works on the rotated QM9. For Cubic2, GPWNO also achieve the state-of-the-art performance among the baseline.
\begin{table*}[h!]
    \centering
    \caption{\textbf{The comparison of NMAE (\%) reported in \citep{cheng2023equivariant}.}}
    \resizebox{1.0\textwidth}{!}{
    \begin{tabular}{llccccccccccc} 
        \toprule
        \multicolumn{2}{l}{Dateset / Model }  & \textbf{Ours} & InfGCN & CNN &
         DeepDFT & DeepDFT2 & EGNN & DimeNet & DimeNet++& GNO & FNO & LNO \\
        \midrule
         \multirow{2}*{QM9} & rot &\textbf{4.68} &4.73  &5.89 &5.87 &4.98 &12.13 &12.98 &12.75 &46.90 &33.25 &24.13 \\
          & unrot &\textbf{0.73} &0.93  &2.01 &2.95 &1.03 &11.92 &11.97 &11.69 &40.86 &28.83 &26.14 \\
        \midrule
        \multirow{6}*{MD}
        &\multicolumn{1}{l}{Ethanol }                &\textbf{4.00} &8.43  &13.97 &7.34 &8.83 &13.90 &13.99 &14.24 &82.35 &31.98 &43.17 \\
        &\multicolumn{1}{l}{Benzene }                &\textbf{2.45} &5.11  &11.98 &6.61 &5.49 &13.49 &14.48 &14.34 &82.46 &20.05 &38.82 \\
        &\multicolumn{1}{l}{Phenol }                 &\textbf{2.68} &5.51  &11.52 &9.09 &7.00 &13.59 &12.93 &12.99 &66.69 &42.98 &60.70 \\
        &\multicolumn{1}{l}{Resorcinol }            &\textbf{2.73} &5.95  &11.07 &8.18 &6.95 &12.61 &12.04 &12.01 &58.75 &26.06 & 35.07 \\
        &\multicolumn{1}{l}{Ethane }              &\textbf{3.67} &7.01  &14.72 &8.31 &6.36 &15.17 &13.11 &12.95 &71.12 &26.31 &77.14 \\
        &\multicolumn{1}{l}{Malonaldehyde }  &\textbf{5.32} &10.34 &18.52 &9.31 &10.68 &12.37 &18.71 &16.79 &84.52 &34.58 &47.22 \\
       \midrule
        \multicolumn{2}{l}{Cubic2 }                &\textbf{7.69} &8.98  & - &14.08 &10.37 &11.74 &12.51 &12.18 &53.55 &48.08 &46.33 \\
        \bottomrule
    \end{tabular}%
    }
    \label{tab:full_table}
\end{table*}

\subsection{Numbers for \cref{fig:Params}}
The numbers for the performance of \cref{fig:Params} is reported in \cref{tab:ablation2}.

 \begin{table}[h!]
    \centering
    \caption{\textbf{Numbers for \cref{fig:Params}.}}
    \begin{tabular}{cccc}
    \toprule
    \# probes $M$ & Benzene & Phenol & \small{Resorcinol} \\
    \midrule
    $10^3$ & 93.81 & 94.06 & 96.11 \\
    $20^3$ & 36.71 & 38.41 & 39.70 \\
    $30^3$ & 14.04 & 15.18 & 15.36 \\
    $40^3$ & 6.92  & 8.27  & 8.12 \\
    $50^3$ & \textbf{4.77}  & \textbf{5.66}  & \textbf{5.18}\\
    \bottomrule 
    \end{tabular}
    \label{tab:ablation2}
\end{table}



\newpage
\section{Complexity Analysis}
\label{sec:complex}

We provide the complexity of our model and the empirical results that demonstrates the analysis.

\textbf{Time complexity.} Our GPWNO scalses linearly with respect to the number of atoms, which is the same order as InfGCN. Our GPWNO scales log linearly with respect to the number of probe nodes due to the Fourier transform, but this is a hyperparameter that one can adjust to the given computational budget.

To be specific, we report the computational complexity of the considered algorithms as follows:
\begin{itemize}
    \item InfGCN\cite{cheng2023equivariant}, DDFT \cite{jorgensen2020deepdft}, DDFT2 \cite{jorgensen2022equivariant}, EGNN \cite{satorras2021n}: $O(d_{\text{atom}}N)$ where $N$ and $d_{\text{atom}}$ denote the number of atoms and maximum degree of the atom graph, respectively.
    \item Ours (GPWNO): $O(d_{\text{atom}}N+d_{\text{probe}}M + M \log M)$ where $M$ and $d_{\text{probe}}$ denotes number of probe nodes and maximum degree of the atom-to-probe node, respectively.
\end{itemize}

Here, the GPWNO complexities $O(d_{\text{atom}}N)$, $O(d_{\text{probe}}M)$, and $O(M\log M)$ are required for atom-wise message passing, atom-to-probe message passing, and Fourier operations, respectively. Since the degree $d_{\text{atom}}$ depends on the cutoff radius, GPWNO scales linearly with the system size, i.e., number of atoms $N$. GPWNO scales log linearly with to number of probe nodes $M$, which is a hyperparameter that one can adjust to the given computational budget.

\begin{figure}
    \centering
    \includegraphics[width=1\linewidth]{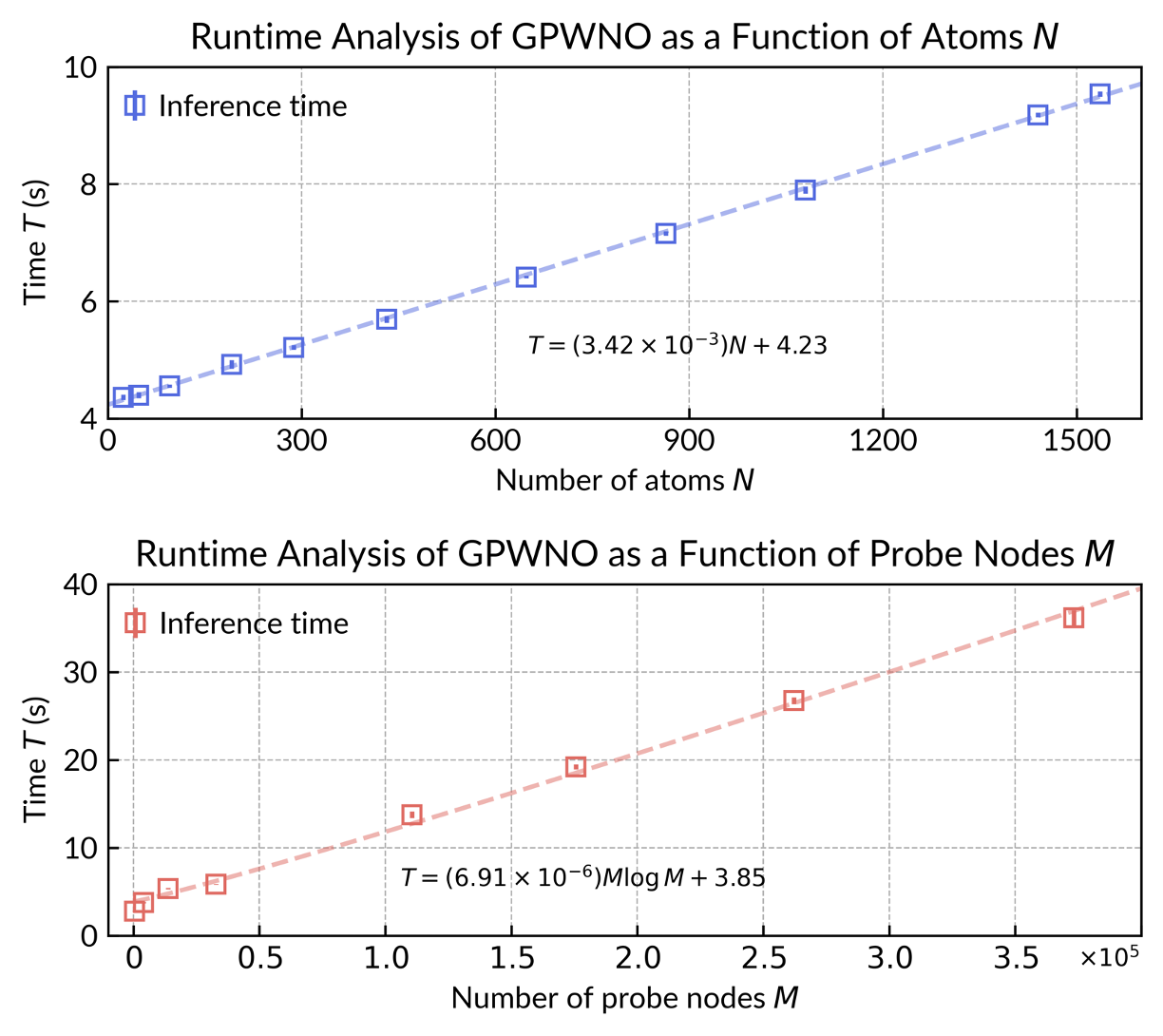}
    \caption{\textbf{Empirical validation of the time complexity by runtime analysis.} The inference times of GPWNO scale linearly with the number of atoms $N$ and log linearly
with the number of probe nodes $M$. }
    \label{fig:complexity}
\end{figure}

\textbf{Empirical result.} We empirically validate how the inference times of GPWNO scale linearly with the number of atoms $N$ and log linearly with the number of probe nodes $M$. To vary $N$, we artificially increase the number of atoms $N$ through supercell expansion, i.e., repeating the same cell to construct a larger supercell. Note that $M$ is a hyperparameter we can tune without depending on the molecule. The results is depicted in \cref{fig:complexity}. These measurements were conducted using 100 forward passes for pre-defined 1000 query points on \ce{SmIn2Cu9} (mp-1209644). \ce{SmIn2Cu9} is sourced from the Materials Project database, consisting of 24 sites per single cell. We report the average and standard deviation over 5 trials. It is worth noting that our GPWNO is much scalable than the traditional DFT solvers, i.e., Linearized coupled cluster doubles method \cite{bartlett1981many,mardirossian2018lowering} has $O(N^5)$ and Kohn-Sham density functional method \cite{kohn1965self} has $O(N^{3})$ complexity.

\end{document}